\documentclass[aps,prd,preprint,nofootinbib,superscriptaddress,11pt]{revtex4-2}

\usepackage[T1]{fontenc}
\usepackage[utf8]{inputenc}
\usepackage{lmodern}
\usepackage{amsmath,amssymb,amsfonts,amsthm,mathtools,mathrsfs,bm}
\usepackage{tikz,graphicx,xcolor,booktabs,multirow}
\usepackage[colorlinks=true,linkcolor=blue,citecolor=blue,urlcolor=blue]{hyperref}

\newcommand{\scri}{\mathscr I}

\providecommand{\eth}{\text{\dh}}
\renewcommand{\eth}{\text{\dh}}

\begin{document}

\title{\textbf{Bondi Four-Momentum Balance in Quantum NSF}\\
\large Physical scattering amplitudes and a numerical energy-balance test}
\author{C.~N.~Kozameh}
\email{carlos.kozameh@unc.edu.ar}
\affiliation{FaMAF, Universidad Nacional de C\'ordoba,
5000 C\'ordoba, Argentina}
\author{G.~O.~Depaola}
\affiliation{FaMAF, Universidad Nacional de C\'ordoba,
5000 C\'ordoba, Argentina}
\date{\today}

\begin{abstract}
The Bondi four-momentum flux-balance laws provide a test of gravitational
scattering in asymptotically flat spacetimes. Starting from the classical
matching conditions at spatial infinity, we formulate their quantum
counterpart in terms of asymptotic radiative observables and the
null-surface formulation (NSF) scattering operator. A principal result
is that the NSF scattering amplitudes already satisfy three of the four
resulting conservation conditions: total spatial momentum is conserved
without imposing momentum-conserving constraints in their dynamical
construction. Energy conservation, by contrast, is not manifest in the NSF
amplitudes and remains an open analytical question. As a limited consistency test of the latter question, we
use the unitary approximation $S_1=\exp(i\delta T_1)$ in the connected
$2\to2$ and $2\to3$ channels, without imposing an energy delta function.
At fixed $E_{\max}=2024E_i$, reducing the infrared cutoff from
$10^{-5}E_i$ to $10^{-8}E_i$ decreases the relative mean-energy residual
by a factor of approximately $584$ and the conditional energy variance
by a factor of approximately $756$. These finite-cutoff trends do not
establish energy conservation for the complete NSF scattering operator;
they motivate an analytical investigation of the remaining condition.
\end{abstract}
\maketitle

\section{Introduction}

The null-surface formulation (NSF) determines gravitational scattering
through a dynamical relation between incoming and outgoing radiative data
at null infinity. In this work we examine the relation between this
scattering map and the Bondi four-momentum flux laws.

The classical analysis builds on the radiative-spacetime framework
developed by Dominguez, Kozameh, and Ludvigsen (DKL)
\cite{DominguezKozamehLudvigsen1996,DominguezKozamehLudvigsen1997}.
Using the matching of the future and past mass aspects at spatial infinity
established by Herberthson and Ludvigsen \cite{HerberthsonLudvigsen1992},
DKL obtained equality of the incoming and outgoing Bondi four-momentum
fluxes in the radiative sector considered here
(Eq.~(14) in Ref.~\cite{DominguezKozamehLudvigsen1996}). The published 1997 paper
\cite{DominguezKozamehLudvigsen1997} develops the associated radiative
phase-space construction. A later treatment of the flux equality was
given by Ashtekar, Campiglia, and Laddha (2018)
\cite{AshtekarCampigliaLaddha2018}, using the Bondi--ADM relation.
Here we formulate the quantum counterpart of the DKL balance relation
and use it as an independent consistency test of the NSF scattering
dynamics. The flux conditions are not inputs to the construction of the
NSF scattering kernels.

The asymptotic Einstein equations constrain the radiative fluxes but
do not determine the outgoing shear from the incoming data. Here,
``dynamics'' refers to this incoming-to-outgoing NSF scattering map,
which supplements the classical balance laws.

A central result is that three of the four translation conservation
conditions are already satisfied by these independently constructed
quantum NSF amplitudes: the kernels make conservation of the three
spatial components manifest, without an externally imposed momentum
delta function. The fourth condition, conservation of total energy,
is not manifest. This nontrivial agreement provides a specific
consistency check of the NSF scattering dynamics, not a proof of
consistency of the complete quantum theory. A limited numerical test
using the retained $2\to2$ and $2\to3$ channels examines whether the
unconstrained dynamics behaves compatibly with energy balance. The test
is auxiliary to the theoretical formulation and cannot establish the
missing conservation law.

The classical review fixes the matching, endpoint, and normalization
conventions needed for this test
\cite{AshtekarHansen1978,HerberthsonLudvigsen1992,Prabhu2019}.
In particular, the endpoint-shear term of the DKL supermomentum has
zero translation projection. Equality of the complete News four-fluxes
therefore does not require vanishing shear memory.

The quantum question requires distinguishing two relations. A unitary
NSF mode map, $a_{\rm out}=S^\dagger a_{\rm in}S$, transports the
quadratic energy observable to $H_{\rm out}=S^\dagger H_{\rm in}S$.
Equality of incoming and outgoing energy is a further condition on
that map. Its classical counterpart is
\begin{equation}
 \mathcal H_+[\mathscr C_{\rm YF}(N_-)]=\mathcal H_-[N_-],
 \label{eq:intro-YF-classical-energy-test}
\end{equation}
where $\mathscr C_{\rm YF}$ is the NSF scattering map obtained by
Yang--Feldman evolution. We formulate the quantum condition through
smeared matrix elements on a common radiative domain
\cite{LSZ1955,Collins2019}. If imposed for arbitrary states, it yields
$(E_f-E_i)\langle f|S|i\rangle=0$.

This distinction leads to two separate questions. The quantum
flux-balance law specifies the energy support that physical scattering
amplitudes must have, independently of how the NSF kernels were obtained.
We may enforce this support on the NSF amplitude as a prescription,
but doing so does not show that it follows from the NSF dynamical
equations. First, we implement this prescription to define a physical
amplitude from the NSF result. Its angular dependence illustrates the
prescription already used in the benchmark sectors of
Parts~I--III \cite{KDI2026,KDII2026,KDIII2026}. The Weinberg comparison
in Part~I uses a Poincar\'e reduction, whose Lorentz covariance permits
evaluation in the center-of-momentum frame. The conservation condition
considered here concerns the BMS translation subgroup. Second, we
remove the added energy constraint and test whether the NSF amplitudes
themselves reproduce the incoming energy in a probability-weighted mean.
The conditional $2\to2$, $2\to3$ calculation tests this question using the retained NSF amplitudes; it is not the full inclusive expectation.

The paper follows this sequence. The classical sections establish the
flux constraint, its endpoint assumptions, and the role of the DKL
translation projection. Section~\ref{sec:quantum-energy-eigenstates}
constructs the quantum flux operators and derives the consequences of
imposing balance, distinguishing them from unitary transport.
Section~\ref{sec:physical-amplitudes} defines the externally constrained
amplitude. Section~\ref{sec:perturbative-scalar-condition} then tests
mean-energy balance without that constraint and examines the infrared
cutoff dependence of the selected channel weights. Deriving the energy
law analytically from the NSF scattering kernels remains the next step.

\subsection{Asymptotic limits and LSZ matrix elements}

The NSF field equations determine null surfaces and radiative data by
propagating information from a spacetime point $x$ to future or past null
infinity.  Passing from this spacetime construction to scattering observables
therefore involves an asymptotic limit.  We interpret the quantum limit in
the weak LSZ sense: for wave-packet-smeared fields (or modes) and physical
scattering states, one takes matrix elements first and then the appropriate
asymptotic limit.  Schematically, if $\mathcal O_x[f]$ denotes the suitably
propagated and smeared bulk observable,
\begin{equation}
 \lim_{x\to\scri^\pm}
 \langle\Phi|\mathcal O_x[f]|\Psi\rangle
 =\langle\Phi|\mathcal O_\pm[f]|\Psi\rangle,
 \label{eq:weak-asymptotic-NSF}
\end{equation}
where the limit is taken along the relevant NSF null geometry.  The limit is defined for smeared matrix elements, as in the LSZ
asymptotic condition
\cite{LSZ1955,Collins2019}.  The analogous weak expression for an
asymptotic balance law, after antipodal matching of labels, is
\begin{equation}
 {}_{\rm out}\!\langle f|
 \bigl(Q^+S^\dagger-S^\dagger Q^-\bigr)|i\rangle_{\rm in}=0.
 \label{eq:intro-weak-balance}
\end{equation}
The domain of states, smearing, and endpoint assumptions are part of this
statement.  A weak relation can nevertheless determine an operator on a
dense domain.  Let $\mathcal D$ be a dense linear subspace of the physical
Hilbert space, contained in the common domain of the relevant asymptotic
operators, and let $A$ denote their difference after both sides have been
represented in the same Hilbert space.  If
\begin{equation}
 \langle\phi|A\psi\rangle=0
 \qquad (\phi,\psi\in\mathcal D),
 \label{eq:weak-to-operator}
\end{equation}
then density implies $A\psi=0$ for every $\psi\in\mathcal D$.
Thus the weak equality gives equality of the operators on $\mathcal D$;
equality of closed operators additionally requires appropriate control of
their domains, for example that $\mathcal D$ is a common core.  If only
diagonal expectation values are known, the same conclusion follows by
polarization when they vanish for every vector in the \emph{linear} space
$\mathcal D$ and define a suitable symmetric quadratic form.  These domain conclusions require the relation for all test vectors;
a restricted family of matrix elements is insufficient. The weak
limit in Eq.~\eqref{eq:weak-asymptotic-NSF} supplies no strong operator
convergence statement.

For the radiative Fock sector used below, one may choose $\mathcal D_{\rm rad}$
to consist of finite-particle wave packets with smooth momentum profiles
supported away from $\omega=0$.  This is a dense linear domain in the usual
radiative Fock space, while excluding a separate zero-frequency soft sector.
Consequently, if the NSF balance relation is established for every pair of
vectors in $\mathcal D_{\rm rad}$, it holds as an operator equality on that
domain.  For the DKL translation charges this radiative representation also
gives the complete Bondi four-momentum: the soft shear term has zero
$\ell=0,1$ projection, as proved in Eq.~\eqref{eq:l-harmonic-property-new}.
This does not establish the analogous statement for general BMS charges,
whose higher harmonics can retain soft and endpoint contributions.

\section{Bondi mass aspect and matching at spatial infinity}

The classical past-to-future flux relation follows the DKL derivation
\cite{DominguezKozamehLudvigsen1996,DominguezKozamehLudvigsen1997}, which uses the Herberthson--Ludvigsen
matching at spatial infinity \cite{HerberthsonLudvigsen1992}. The
Ashtekar--Hansen framework \cite{AshtekarHansen1978} supplies the
geometric setting. The Hamiltonian interpretation of radiative fluxes
on a null boundary follows Ashtekar and Streubel
\cite{AshtekarStreubel1981}. Equality of the complete incoming and
outgoing fluxes additionally uses the matching between $\scri^-$ and
$\scri^+$ and the stated timelike endpoint conditions, as in DKL.

For comparison, Ashtekar, Campiglia, and Laddha
\cite{AshtekarCampigliaLaddha2018}, Sec.~V, discuss the same
four-momentum flux equality using the Bondi--ADM relation of
Ref.~\cite{AshtekarMagnonAshtekar1979}. We recall the matching and
endpoint assumptions before extracting the translation components of
the DKL relation. The zero translation projection of the endpoint-shear
term permits nonzero memory; the flux itself still depends on the
radiative data.

\subsection{Definition of the Bondi mass aspect}

On future and past null infinity we use the Bondi mass aspect
\begin{equation}
 m_B^\pm
 =-\operatorname{Re}\!\left(
 \Psi_2^{0\pm}+\sigma^{0\pm}\dot{\bar\sigma}^{0\pm}
 \right).
 \label{eq:Bondi-mass-aspect-new}
\end{equation}
Here $\Psi_2^{0\pm}$ denotes the leading Coulombic Weyl coefficient, $\sigma^{0\pm}$ the asymptotic shear, and the dot denotes differentiation with respect to the appropriate null time: retarded time $u$ on $\scri^+$ and advanced time $v$ on $\scri^-$.  The News is $N^\pm=\dot\sigma^{0\pm}$.

At the endpoints adjacent to spatial infinity the News vanishes,
\begin{equation}
 \left.\dot\sigma^{0+}\right|_{\scri^+_-}=0,
 \qquad
 \left.\dot\sigma^{0-}\right|_{\scri^-_+}=0.
 \label{eq:news-zero-i0-new}
\end{equation}
The shear itself need not vanish.  Nevertheless,
\begin{equation}
 \sigma^{0\pm}\dot{\bar\sigma}^{0\pm}\big|_{i^0}=0,
\end{equation}
and therefore
\begin{equation}
 m_B^\pm(i^0,\Omega)
 =-\operatorname{Re}\Psi_2^{0\pm}(i^0,\Omega).
 \label{eq:mass-aspect-Weyl-i0-new}
\end{equation}

\subsection{Antipodal continuity through \texorpdfstring{$i^0$}{spatial infinity}}

We restrict the classical discussion to radiative spacetimes for
which the matching hypotheses at spatial infinity hold. In addition
to an Ashtekar--Hansen completion, the rescaled real Coulombic Weyl
component must have finite, compatible limits from the null and
spatial directions in matched Bondi frames. The shear has finite
endpoint limits, the News tends to zero, and its decay is sufficient
for the integrated fluxes below to exist. These are regularity and
matching hypotheses on the solution, not consequences of the
condition \(N^\pm(i^0)=0\) alone. They are the assumptions needed
when applying the Weyl matching of
Ref.~\cite{HerberthsonLudvigsen1992}; a precise sufficient
null-regularity framework for comparing the null and spatial limits
is given in Ref.~\cite{Prabhu2019}, Definition~4.1.

Ashtekar--Hansen provide the common structure at spatial infinity,
while Herberthson and Ludvigsen construct matched Bondi scalings and
obtain the corresponding Weyl-curvature matching. In those frames,
with the standard antipodal identification, this gives
\begin{equation}
 \Psi_2^{0+}(i^0,\Omega)
 =\Psi_2^{0-}(i^0,\mathcal A\Omega)
 \label{eq:Weyl-antipodal-new}
\end{equation}
for the relevant real Coulombic component, with $\mathcal A$ the antipodal map.  Using Eq.~\eqref{eq:mass-aspect-Weyl-i0-new}, the matching extends directly to the Bondi mass aspect,
\begin{equation}
 m_B^+(i^0,\Omega)
 =m_B^-(i^0,\mathcal A\Omega).
 \label{eq:mass-aspect-antipodal-new}
\end{equation}
The vanishing News term in Eq.~\eqref{eq:Bondi-mass-aspect-new}
relates Weyl matching to mass-aspect continuity at $i^0$.

\section{The DKL supermomentum function \texorpdfstring{$\chi$}{chi}}

The DKL representation expresses the matched endpoint charge in
terms of the shear and its News. This identifies which part of the
asymptotic constraint is measured by radiative gravitons and which
part belongs to the endpoint shear.

\subsection{Definition from the mass aspect}

In the radiative phase space of Dominguez, Kozameh, and Ludvigsen, the geometry is flat sufficiently near timelike infinity.  On $\scri^+$ one has $\Psi_2^{0+}\to0$ as $u\to+\infty$, while the nontrivial Coulombic information at the opposite endpoint is encoded in the spin-weight-zero function
\begin{equation}
 \chi^+(\Omega)
 \equiv -\lim_{u\to-\infty}\operatorname{Re}\Psi_2^{0+}(u,\Omega)
 =m_B^+(i^0,\Omega).
 \label{eq:chi-plus-definition}
\end{equation}
The corresponding function on $\scri^-$ is
\begin{equation}
 \chi^-(\Omega)
 \equiv -\lim_{v\to+\infty}\operatorname{Re}\Psi_2^{0-}(v,\Omega)
 =m_B^-(i^0,\Omega).
 \label{eq:chi-minus-definition}
\end{equation}
The equalities with the mass aspect again use $N^\pm(i^0)=0$.

Equations~\eqref{eq:chi-plus-definition} and \eqref{eq:chi-minus-definition}
define $\chi^+$ and $\chi^-$ on the two null boundaries. Their relation
follows from antipodal continuity, Eq.~\eqref{eq:mass-aspect-antipodal-new}.

\subsection{DKL flux formula}

Following DKL \cite{DominguezKozamehLudvigsen1997}, the asymptotic
shear is defined in the Bondi tetrad by
\begin{equation}
 \sigma^\pm=m_\pm^a m_\pm^b\nabla_a\ell_b^\pm.
 \label{eq:DKL-shear-tetrad-definition}
\end{equation}
Under $m_\pm^a\mapsto e^{i\lambda}m_\pm^a$ it transforms as
$\sigma^\pm\mapsto e^{2i\lambda}\sigma^\pm$. Thus $\sigma^\pm$
has spin weight $+2$ on both null boundaries, while its complex
conjugate has spin weight $-2$. The scalar angular term is therefore
$\bar\eth_\pm^{\,2}\sigma_0^\pm$, or its complex conjugate
$\eth_\pm^{\,2}\bar\sigma_0^\pm$. The angular dyads on the two
boundaries are compared using the antipodal identification.

In the DKL endpoint frame the shear vanishes at timelike infinity,
and $\sigma_0^\pm$ is its limit at spatial infinity. To allow a
nonzero constant endpoint shear, we use the corresponding differences,
with the retarded and advanced times of the present paper:
\begin{equation}
 \sigma_0^-(\Omega)
 =\sigma^-(i^0,\Omega)-\sigma^-(i^-,\Omega).
 \label{eq:sigma0minus-new}
\end{equation}
Let
\begin{equation}
 \sigma_0^+(\Omega)
 =\sigma^+(i^0,\Omega)-\sigma^+(i^+,\Omega)
 \label{eq:sigma0plus-new}
\end{equation}
be the future infrared shear difference.  The Bondi evolution equation integrated along all of $\scri^+$ yields the DKL relation
\begin{equation}
 \chi^+(\Omega)
 =\operatorname{Re}\!\left[\bar\eth^{\,2}\sigma_0^+(\Omega)\right]
 +\int_{-\infty}^{+\infty}du\,
 \dot\sigma^+(u,\Omega)\dot{\bar\sigma}^{+}(u,\Omega).
 \label{eq:DKL-chi-plus}
\end{equation}
The analogous relation on $\scri^-$ will be written as
\begin{equation}
 \chi^-(\Omega)
 =\operatorname{Re}\!\left[\bar\eth_-^{\,2}\sigma_0^-(\Omega)\right]
 +\int_{-\infty}^{+\infty}dv\,
 \dot\sigma^-(v,\Omega)\dot{\bar\sigma}^{-}(v,\Omega),
 \label{eq:DKL-chi-minus}
\end{equation}
The operators $\eth_-$ and $\bar\eth_-$ refer to the past Bondi
chart, with the antipodal identification understood. The scalar shear
term can equivalently be written as
$\operatorname{Re}[\eth_-^{\,2}\bar\sigma_0^-]$.
Thus $\chi^\pm$ contains the endpoint-shear term and the integrated
positive News flux density.

On $\scri^+$, this decomposition is
\begin{equation}
 \chi^+=\chi_{\rm IR}^+ + \chi_{\rm rad}^+,
 \qquad
 \chi_{\rm IR}^+=\operatorname{Re}\!\left[\bar\eth^{\,2}\sigma_0^+\right],
 \qquad
 \chi_{\rm rad}^+=\int_{-\infty}^{+\infty}du\,
 \dot\sigma^+\dot{\bar\sigma}^{+}.
 \label{eq:chi-IR-rad-split}
\end{equation}
Under a pure BMS supertranslation, both terms are invariant at the
matched endpoints.  In a fixed Bondi conformal frame, a
supertranslation by an angle-dependent function $f(\Omega)$ shifts the
null time by the same amount at both endpoints of each generator and
adds the same time-independent angular term to the shear at both ends.
Hence the difference $\sigma_0^+$ in
Eq.~\eqref{eq:sigma0plus-new} is unchanged when the endpoints are
compared geometrically.  The News transforms by relabeling its null
time \cite{Ashtekar2015,Strominger2014}, so its complete integral along
each generator also satisfies
\begin{equation}
 \int_{-\infty}^{+\infty}du'\,|N'(u',\Omega)|^2
 =
 \int_{-\infty}^{+\infty}du\,|N(u,\Omega)|^2.
 \label{eq:news-flux-supertranslation-invariance}
\end{equation}
Both terms in Eq.~\eqref{eq:chi-IR-rad-split}, and therefore
$\chi^+(i^0,\Omega)$, are invariant under this pure
supertranslation, subject to the stated endpoint limits.  The same
reasoning applies on $\scri^-$, with antipodally matched
supertranslation parameters at $i^0$.  The invariance concerns the endpoint function $\chi$ and the full
flux integral; the mass aspect at finite null time has a different
transformation law.

At future timelike infinity we assume $\Psi_2^{0+}\to0$ and
$\dot\sigma^{0+}\to0$, with finite endpoint shear. By
Eq.~\eqref{eq:Bondi-mass-aspect-new}, these conditions imply
$m_B^+\to0$. With angular convergence sufficient to pass the limit
through the charge integral, all four translation components therefore
vanish:
\begin{equation}
 P_B^{+a}(i^+)
 =\frac{1}{4\pi G}\lim_{u\to+\infty}
 \int_{S^2}d\Omega\,l^a(\Omega)m_B^+(u,\Omega)=0.
 \label{eq:timelike-endpoint-fourmomentum-zero}
\end{equation}
The analogous past endpoint conditions give $P_B^{-a}(i^-)=0$.
Thus vanishing endpoint four-momentum follows from the assumed
asymptotic fields and the definition of the Bondi charge. It does not
follow merely from vanishing News or from a choice of center-of-momentum
frame. These conditions specify the completely dispersive radiative
sector and exclude a massive recoiling remnant; they permit nonzero
Bondi four-momentum at spatial infinity carried by the incoming and
outgoing radiation.

The DKL condition $m_B(i^\pm)=0$ allows nonzero endpoint shear:
$\dot\sigma(i^\pm)=0$ and $\Psi_2^0(i^\pm)=0$ do not require
$\sigma(i^\pm)=0$. Nonzero memory gives a distributional $1/\omega$
term in the shear transform, as detailed in
Appendix~\ref{app:news-memory}. Its translation projection is evaluated
below.

\subsection{Conservation of \texorpdfstring{$\chi$}{chi} across spatial infinity}

Equations~\eqref{eq:chi-plus-definition}, \eqref{eq:chi-minus-definition}, and \eqref{eq:mass-aspect-antipodal-new} give
\begin{equation}
 \chi^+(\Omega)=\chi^-(\mathcal A\Omega).
 \label{eq:chi-antipodal-matching}
\end{equation}
Thus the DKL supermomentum leaf label is the same physical datum on the two sides of spatial infinity, after the standard antipodal identification.  Its translation projection gives the matching of Bondi four-momentum.
This is a classical statement under the matching and endpoint
hypotheses. The foliation by fixed $\chi$ supplies the starting point
for the sectorwise construction of Ref.~\cite{DominguezKozamehLudvigsen1997};
it does not require the radiative quantization below to be performed
on an individual leaf.

\section{Bondi four-momentum from the translation components of \texorpdfstring{$\chi$}{chi}}

We now extract the four translation charges from the angular DKL
relation. This projection removes the endpoint-shear term and provides
the News-flux observables that will be quantized below. It does not
require choosing a preferred Lorentz subgroup of BMS.

The four translation components are obtained by projecting the supermomentum function onto the $\ell=0,1$ harmonics.  With
\begin{equation}
 l^a(\Omega)=\bigl(1,\widehat{\mathbf k}(\Omega)\bigr),
 \label{eq:l-a-new}
\end{equation}
we define
\begin{equation}
 P_B^{\pm a}(i^0)
 =\frac{1}{4\pi G}\int_{S^2}d\Omega\,
 l^a(\Omega)\chi^\pm(\Omega),
 \label{eq:Bondi-P-from-chi}
\end{equation}
The mass aspect retains the Newman--Penrose normalization in
Eq.~\eqref{eq:Bondi-mass-aspect-new}, with \(l^0=1\) and the unit-sphere
area element. Thus a nonradiating Schwarzschild check,
\(\Psi_2^0=-GM\) and \(\sigma^0=0\), gives \(P_B^0=M\).
This fixes the charge normalization independently of the oscillator
conventions. The angular labels on \(\scri^-\) are understood
antipodally when comparing physical four-vectors.

\subsection{Lorentz covariance at spatial infinity}
\label{subsec:lorentz-covariance-i0}

The Lorentz transformation of Eq.~\eqref{eq:Bondi-P-from-chi} follows
from the conformal weights of its factors.  Write the conformal action of a proper
Lorentz transformation on the celestial sphere as
$\Omega'=g(\Omega)$, with conformal factor $K(\Omega)>0$ chosen so that
\begin{equation}
 d\Omega'=K^2d\Omega,\qquad
 l'^{a}(\Omega')=K\Lambda^a{}_b l^b(\Omega).
 \label{eq:lorentz-sphere-transform}
\end{equation}
The second relation follows by normalizing the transformed future null
vector to $l'^0=1$.  At the stationary endpoint $i^0$, where the News
and its endpoint derivatives vanish, the BMS transformation law of the
mass aspect has conformal weight minus three in these conventions:
\begin{equation}
 \chi'(i^0,\Omega')=K^{-3}\chi(i^0,\Omega).
 \label{eq:chi-lorentz-weight}
\end{equation}
This is the endpoint form of the Bondi mass-aspect transformation; the
additional News-dependent terms at finite null time are absent under
the regularity assumptions used here
\cite{Ashtekar2015,ChenWangWangYau2023}.
Consequently,
\begin{align}
 P_B'{}^a(i^0)
 &=\frac{1}{4\pi G}\int_{S^2}d\Omega'\,l'^a(\Omega')\chi'(i^0,\Omega')\nonumber\\
 &=\frac{1}{4\pi G}\int_{S^2}K^2d\Omega\,
       (K\Lambda^a{}_b l^b)(K^{-3}\chi)\nonumber\\
 &=\Lambda^a{}_bP_B^b(i^0).
 \label{eq:Bondi-P-Lorentz-covariance}
\end{align}
In particular, a rotation rotates the spatial momentum without
changing the energy, while a boost mixes the energy and the
longitudinal momentum.  These are the transformation properties of the translation charges.
The rotation and boost charges $J^{ab}$ are distinct observables.

The relevant BMS group relations are
\begin{equation}
 \mathcal T\subset\mathcal{ST}\subset\mathrm{BMS}_4,
 \qquad
 \mathrm{BMS}_4/\mathcal{ST}\simeq SO^+(3,1),
 \label{eq:BMS-translation-subgroup-quotient}
\end{equation}
where $\mathcal{ST}$ is the Abelian normal subgroup of supertranslations
and $\mathcal T$ is the four-dimensional normal subgroup of ordinary
translations \cite{Ashtekar2015}.  In a Bondi conformal frame the
translation parameters satisfy $\eth^2\alpha=0$ and contain only
$\ell=0,1$ harmonics.  Since supertranslations commute with translations,
the conjugation action of BMS on $\mathcal T$ descends to the Lorentz
quotient.  This action maps a translation four-vector to another
translation four-vector.  Equation~\eqref{eq:Bondi-P-Lorentz-covariance}
checks that the corresponding Bondi charges transform as
$P_B'{}^a=\Lambda^a{}_bP_B^b$.  No choice of a preferred full
Poincar\'e subgroup is required for this translation-sector statement.

The pure infrared shear term in $\chi$ does not contribute to these four translation components.  Indeed, using integration by parts on the sphere,
\begin{align}
 \int_{S^2}d\Omega\,l^a\,
 \operatorname{Re}\!\left[\bar\eth^{\,2}\sigma_0^+\right]
 &=\operatorname{Re}\!\int_{S^2}d\Omega\,\sigma_0^+\,\eth^2 l^a=0,
 \end{align}
with the analogous result on $\scri^-$.  The last equality follows because $l^a$ contains only the $\ell=0,1$ harmonics, so that
\begin{equation}
 \eth^2l^a=\bar\eth^{\,2}l^a=0.
 \label{eq:l-harmonic-property-new}
\end{equation}
Consequently, although the full supermomentum $\chi$ contains both infrared and radiative sectors, the Bondi four-momentum obtained from its translation components is insensitive to the infrared term.  Substituting Eq.~\eqref{eq:DKL-chi-plus} into Eq.~\eqref{eq:Bondi-P-from-chi} gives explicitly
\begin{equation}
 P_B^{+a}(i^0)
 =\frac{1}{4\pi G}\int_{S^2}d\Omega\,l^a(\Omega)
 \int_{-\infty}^{+\infty}du\,
 \dot\sigma^+(u,\Omega)\dot{\bar\sigma}^{+}(u,\Omega),
 \label{eq:Bondi-P-radiative-only}
\end{equation}
with the corresponding expression on $\scri^-$.  The translation balance therefore depends only on the News flux,
although the DKL supermomentum retains the soft-sector information. Thus the same four-momentum is obtained whether one uses the DKL representative or the corresponding Tamburino--Winicour/Ashtekar representative of the supermomentum.

Using Eq.~\eqref{eq:chi-antipodal-matching} together with the standard antipodal identification of physical null directions gives
\begin{equation}
 P_B^{-a}(i^0)=P_B^{+a}(i^0).
 \label{eq:four-momentum-matching-new}
\end{equation}
This translation projection does not determine the infrared sector.
For antipodally identified data, write
$q^\pm:=\chi_{\rm IR}^\pm$ and $F^\pm:=\chi_{\rm rad}^\pm$.
The matching $\chi^+=\chi^-$ gives
\[
 q^+-q^-=-(F^+-F^-),\qquad
 \int d\Omega\,l^a q^\pm=0.
\]
Consequently, the incoming and outgoing integrated News four-fluxes
can agree even when $q^+\ne q^-$. Higher angular components of the
News flux can compensate a change of the soft contribution. The four
translation charges therefore do not determine whether scattering
preserves the infrared sector. This is the translation-sector
consequence of the soft--hard charge decomposition in asymptotic
quantization \cite{AshtekarCampigliaLaddha2018}. Equality of the fluxes
uses the matching and endpoint conditions; it is not a statement that
arbitrary changes of radiative data leave the flux unchanged.

For the translation sector this common value agrees with the ADM four-momentum under the usual regularity assumptions at spatial infinity,
\begin{equation}
 P_B^{-a}(i^0)=P_{\rm ADM}^a=P_B^{+a}(i^0).
 \label{eq:Bondi-ADM-new}
\end{equation}

\section{Bianchi flux laws on \texorpdfstring{$\scri^-$ and $\scri^+$}{past and future null infinity}}

The Bondi four-momentum and the News flux are distinct quantities.  Define the positive integrated News four-fluxes by
\begin{equation}
 \mathcal F_N^{\pm a}
 =\frac{1}{4\pi G}
 \int_{\scri^\pm}d\tau_\pm\,d\Omega\,
 \widehat\ell_\pm^a(\Omega)
 N^\pm(\tau_\pm,\Omega)\overline N^\pm(\tau_\pm,\Omega),
 \label{eq:classical-News-four-flux-new}
\end{equation}
with $\tau_+=u$ and $\tau_-=v$. The charge directions are normalized
by \(\widehat\ell_\pm^0=1\), as in Eq.~\eqref{eq:l-a-new};
the null directions in Part~I cut kernels can instead carry the
explicit \(1/\sqrt2\) tetrad normalization used below.
The time components $\mathcal F_N^{\pm0}$ are non-negative.
The prefactor uses the same Newman--Penrose shear as the mass
aspect. For a unit-sphere dyad with \(m_A\bar m^A=1\), the tensor
convention
\[
 \sigma_{AB}=2\bigl(\bar\sigma\,m_Am_B+
                   \sigma\,\bar m_A\bar m_B\bigr)
\]
gives \(N_{AB}N^{AB}=8|\dot\sigma|^2\). The tensor flux
\((32\pi G)^{-1}\int N_{AB}N^{AB}\) therefore equals the scalar
flux in Eq.~\eqref{eq:classical-News-four-flux-new}, with scalar
prefactor $1/(4\pi G)$.
With the Fourier convention
$\widetilde N^\pm(\omega,\Omega)
 =\int_{-\infty}^{+\infty}d\tau_\pm\,
 e^{-i\omega\tau_\pm}N^\pm(\tau_\pm,\Omega)$,
Parseval's identity at each angle gives
\begin{equation}
 \int_{-\infty}^{+\infty}d\tau_\pm\,
 |N^\pm(\tau_\pm,\Omega)|^2
 =\int_{-\infty}^{+\infty}\frac{d\omega}{2\pi}\,
 |\widetilde N^\pm(\omega,\Omega)|^2.
 \label{eq:classical-News-Parseval}
\end{equation}
Substitution into Eq.~\eqref{eq:classical-News-four-flux-new}
rewrites the same classical flux as
\begin{equation}
 \mathcal F_N^{\pm a}
 =\frac{1}{4\pi G}
 \int d\Omega\,\widehat\ell_\pm^a(\Omega)
 \int_{-\infty}^{+\infty}\frac{d\omega}{2\pi}\,
 \bigl|\widetilde N^\pm(\omega,\Omega)\bigr|^2.
 \label{eq:classical-News-four-flux-Fourier}
\end{equation}
Quantization of the radiative Fourier modes in
Sec.~\ref{sec:quantum-energy-eigenstates} gives the normal-ordered
operators $P_\pm^a$ and their temporal components,
$H_{\rm in}=P_-^0$ and $H_{\rm out}=P_+^0$. These quantum observables
are constructed after deriving the classical flux balance.

On past null infinity, incoming radiation increases the Bondi four-momentum from $i^-$ to $i^0$,
\begin{equation}
 P_B^{-a}(i^0)-P_B^{-a}(i^-)=\mathcal F_N^{-a}.
 \label{eq:past-balance-new}
\end{equation}
In particular,
\begin{equation}
 E_B(i^0)=E_B(i^-)+\mathcal F_N^{-0}
 \ge E_B(i^-).
 \label{eq:past-energy-new}
\end{equation}

On future null infinity, outgoing radiation decreases the Bondi four-momentum,
\begin{equation}
 P_B^{+a}(i^+)-P_B^{+a}(i^0)=-\mathcal F_N^{+a}.
 \label{eq:future-balance-new}
\end{equation}
Hence
\begin{equation}
 E_B(i^+)=E_B(i^0)-\mathcal F_N^{+0}
 \le E_B(i^0).
 \label{eq:future-energy-new}
\end{equation}
\section{Global classical Bondi four-momentum balance}

Eliminating the common four-momentum at $i^0$ from
Eqs.~\eqref{eq:past-balance-new} and \eqref{eq:future-balance-new}
using Eq.~\eqref{eq:four-momentum-matching-new} gives
\begin{equation}
 P_B^{-a}(i^-)+\mathcal F_N^{-a}
 =P_B^{+a}(i^+)+\mathcal F_N^{+a}.
 \label{eq:global-classical-balance-new}
\end{equation}
Equation~\eqref{eq:global-classical-balance-new} relates the endpoint
Bondi four-momenta to the complete News fluxes. In general, conservation
applies to their sum, rather than to either contribution separately.

\subsection{DKL radiative sector}

For the DKL radiative spacetimes there are no horizons or massive remnants and the geometry becomes flat at the temporal endpoints.  The endpoint Bondi four-momenta vanish,
\begin{equation}
 P_B^{-a}(i^-)=0,
 \qquad
 P_B^{+a}(i^+)=0.
 \label{eq:DKL-endpoints-new}
\end{equation}
Then the general classical law reduces to
\begin{equation}
 \mathcal F_N^{-a}=\mathcal F_N^{+a}.
 \label{eq:DKL-classical-flux-equality-new}
\end{equation}
This is the DKL conservation of the total Bondi momentum flux
\cite{DominguezKozamehLudvigsen1996};
Eq.~(14) of the preprint states it explicitly for translation parameters.
The quantum flux operators are constructed in
Sec.~\ref{sec:quantum-energy-eigenstates}. Extending the classical
equality to quantum matrix elements is an additional working assumption.

\subsection{Energy-flux conservation as a dynamical condition on NSF}
\label{subsec:classical-NSF-flux-condition}

Equation~\eqref{eq:DKL-classical-flux-equality-new} must also be
implemented as a global dynamical consistency condition on the classical
NSF scattering solution. Denote its outgoing shear by
\begin{equation}
 \sigma^+=\mathscr C[\sigma^-],
 \qquad
 \mathcal H_\pm[\sigma^\pm]
 :=\frac{1}{4\pi G}\int_{\scri^\pm}d\tau_\pm\,d\Omega\,
 |\partial_{\tau_\pm}\sigma^\pm|^2.
 \label{eq:NSF-classical-energy-functionals}
\end{equation}
Dependence of the map on the complex conjugate shear is implicit.
The normalization is that of
Eq.~\eqref{eq:classical-News-four-flux-new}. With the matching and
vanishing endpoint charges specified above, the required equation is
\begin{equation}
 \boxed{\mathcal H_+\bigl[\mathscr C[\sigma^-]\bigr]
             -\mathcal H_-[\sigma^-]=0.}
 \label{eq:NSF-classical-dynamical-energy-condition}
\end{equation}
It restricts the outgoing data produced by the NSF field equations
for each admissible incoming configuration. Defining a map between
the two boundaries does not by itself establish this equality.
We therefore include Eq.~\eqref{eq:NSF-classical-dynamical-energy-condition}
explicitly among the conditions on the NSF scattering problem. It is
a consequence of classical gravitational dynamics with the stated
boundary conditions, rather than a new local field equation. Our
perturbative construction has not previously used it as an explicit
constraint on the recursively computed shear corrections; whether
those corrections already satisfy it must be checked.

To formulate the check without a perturbative truncation, absorb the
free antipodal and spin-frame identifications into the radiative data.
On the common null-time and angular labels, define the full nonlinear
correction by
\begin{equation}
 N_{\rm out}=N_{\rm in}+\Delta N,
 \qquad
 \Delta N=\partial_\tau(\sigma_{\rm out}-\sigma_{\rm in}).
 \label{eq:NSF-classical-News-full-correction}
\end{equation}
The free identification preserves the integrated News norm. Thus
Eq.~\eqref{eq:NSF-classical-dynamical-energy-condition} is equivalent to
\begin{equation}
 \boxed{\frac{1}{4\pi G}\int d\tau\,d\Omega\,
 \left[2\operatorname{Re}\!\left(\overline{N_{\rm in}}\Delta N\right)
       +|\Delta N|^2\right]=0.}
 \label{eq:NSF-classical-flux-exact-check}
\end{equation}
This is the equation to be verified for the full nonlinear correction
provided by the classical NSF dynamical map, for each admissible incoming datum.
The interference and quadratic terms enter together; neither is required
to vanish separately. No expansion in a bookkeeping parameter or
truncation is used in this condition. Establishing it directly from the
NSF field equations is left for future work. Its quantum counterpart is
the additional flux balance condition developed below, distinct from
unitary transport of the flux observable.

\section{Quantum four-momentum balance in the radiative sector}
\label{sec:quantum-energy-eigenstates}

The classical result specifies the flux that a scattering solution
must preserve. We now ask how this requirement is represented for
quantum radiative states. The construction has two ingredients: the
NSF map determines the outgoing operators, while the classical balance
law motivates a condition on their fluxes. Keeping these steps distinct
allows us subsequently to test the first without imposing the second.

\subsection{Review of the NSF quantum setup and its assumptions}

The linearized NSF map identifies the free incoming and outgoing
radiative data by the antipodal direction and spin-frame map
$\mathcal A_{\rm rad}$. Absorbing it into the mode labels makes the
free map the identity. For the nonlinear quantum construction we use
the following assumptions:
\begin{enumerate}
\item The asymptotic Hilbert spaces $\mathcal H_\pm$ admit compatible
unitary identifications $J_\pm$ with a common radiative Fock space
$\mathcal F_{\rm rad}$. A unitary state map
$\mathcal U:\mathcal H_-\to\mathcal H_+$ is represented, in our
adjoint convention, by
\[
 S^\dagger=J_+\mathcal U J_-^\dagger,
 \qquad a_{\rm out}=S^\dagger a_{\rm in}S.
\]
The represented vacuum is invariant up to a phase. These are
implementation assumptions; unitarity alone does not impose energy
balance.
\item We write $S=e^{iT}$ with a self-adjoint phase. Inverse BCH
reconstructs its formal coefficients from the NSF mode variations;
it does not establish convergence of the infinite series or, by
itself, a self-adjoint realization.
\item The fluxes and their products are defined on the common dense
radiative domain specified below. Whenever multiplication by $S$ or
$S^\dagger$ is used, this domain is assumed invariant. Sharp modes
stand for smeared or regulated matrix elements, as in
Eq.~\eqref{eq:weak-asymptotic-NSF}.
\item The normal-ordered flux operators quantize the classical News
functionals in the chosen finite-frequency representation. We assume
neither a representation of all infrared data nor preservation of an
infrared sector. Equality of incoming and outgoing quantum fluxes is
the additional correspondence condition examined below.
\end{enumerate}

\subsection{Inverse BCH reconstruction of the NSF generators}
\label{subsec:inverse-BCH-summary}

Quantizing the classical NSF scattering map gives the outgoing mode
variations in terms of incoming operators
\cite{KDI2026,KDII2026,KDIII2026}. Inverse BCH reconstruction associates
these variations with Hermitian generators $\delta T_n$:
\begin{equation}
 S=e^{iT},\qquad
 T=\varepsilon\delta T_1+\varepsilon^2\delta T_2+\cdots,
 \qquad a_{\rm out}=e^{-iT}a_{\rm in}e^{iT}.
 \label{eq:BCH-unitary-reconstruction}
\end{equation}
Writing
$a_{\rm out}=a_{\rm in}+\varepsilon\delta a_2
+\varepsilon^2\delta a_3+\cdots$, expansion of the adjoint action gives
\begin{equation}
 \begin{aligned}
 \delta a_2&=i[a_{\rm in},\delta T_1],\\
 \delta a_3&=i[a_{\rm in},\delta T_2]
 -\frac12[[a_{\rm in},\delta T_1],\delta T_1].
 \end{aligned}
 \label{eq:BCH-first-reconstruction-equations}
\end{equation}
Thus the known lower-order nested commutator is subtracted from
$\delta a_3$ to obtain the equation for $\delta T_2$. The parameter
$\varepsilon$ records formal order and is set to one after organizing
the contributions; it supplies no numerical suppression.

The CCR compatibility conditions and the reconstruction proof are
given in Ref.~\cite{KozamehBCH2026}. Here we use the resulting generators,
with the implementation and domain assumptions stated above, rather
than repeat that construction. The phase $T$ is distinct from the
Bondi energy operator: its reconstruction does not impose energy-flux
balance. The generators enter the amplitude expansion used below.

\subsection{Radiative representation and flux operators}

Equation~\eqref{eq:l-harmonic-property-new} permits the translation
fluxes to be expressed through radiative modes. Their classical
functional also applies to data with memory, although the chosen
Fock representation does not describe the complete infrared data
\cite{Ashtekar2015}.

We take $\mathcal D_{\rm rad}$ to be a common dense radiative domain,
with finite-particle states smeared by smooth mode functions of compact
support in $0<\omega<\infty$ as test states. Each test function is supported away from zero, with no common
positive-frequency cutoff required for the domain. The flux operators, scattering
map and compositions used below are assumed to be defined on the
stated domain. Sharp modes serve as distributional notation for their
regulated matrix elements. For a radiative mode,

$k^a=\omega_k l^a(\Omega_k)$, where
$\omega_k=|\mathbf k|$ and $l^a=(1,\widehat{\mathbf k})$ as in
Eq.~\eqref{eq:l-a-new}.  Quantizing the radiative Fourier modes of the classical flux
\eqref{eq:classical-News-four-flux-Fourier} and normal ordering gives, in the Part~I oscillator normalization,
\begin{equation}
 \begin{aligned}
 P_-^a\equiv\widehat{\mathcal F}_N^{-a}
 &=2\sum_\lambda\int d\mu(k)\,\omega_k l^a(\Omega_k)
 a_{{\rm in},\lambda}^\dagger(k)a_{{\rm in},\lambda}(k),\\
 P_+^a\equiv\widehat{\mathcal F}_N^{+a}
 &=2\sum_\lambda\int d\mu(k)\,\omega_k l^a(\Omega_k)
 a_{{\rm out},\lambda}^\dagger(k)a_{{\rm out},\lambda}(k).
 \end{aligned}
 \label{eq:LSZ-H-check}
\end{equation}
We use the Part~I normalization, with \(c=\hbar=1\) and \(G\)
explicit:
\begin{equation}
 [a_\lambda(\vec k),a_{\lambda'}^\dagger(\vec k')]
 =\omega_k\delta_{\lambda\lambda'}\delta^{(3)}(\vec k-\vec k')
,
 \qquad d\mu(k)=\frac{d^3k}{2\omega_k}.
 \label{eq:QS-I-CCR-normalization}
\end{equation}
The factor $2$ in the charges
ensures $[P_-^a,a_\lambda^\dagger(k)]=k^a a_\lambda^\dagger(k)$.
In spherical momentum coordinates, with
\(a_\lambda(\omega,\Omega):=a_\lambda(\omega\widehat{\mathbf k})\),
the CCR become
\begin{equation}
 [a_\lambda(\omega,\Omega),a_{\lambda'}^\dagger(\omega',\Omega')]
 =\frac1{\omega}\delta_{\lambda\lambda'}\delta(\omega-\omega')
   \delta_{S^2}(\Omega,\Omega').
 \label{eq:spherical-mode-CCR-conversion}
\end{equation}
Here \(a_\lambda(\omega,\Omega)\) is the momentum-normalized
operator evaluated at \(\vec k=\omega\widehat{\mathbf k}\).
The physical NP shear corresponding to this convention is
\begin{equation}
 \sigma(u,\Omega)=\sqrt{2G}\int_0^\infty d\omega\,
 \left[
 a_+(\omega,\Omega)e^{-i\omega u}
 +a_-^\dagger(\omega,\Omega)e^{i\omega u}
 \right],
 \label{eq:NP-shear-momentum-normalized}
\end{equation}
with the appropriate incoming or outgoing mode family on each
boundary. This equation fixes the conversion of shear Fourier
coefficients to the momentum convention used here; Fourier
coefficients normalized with \(d\omega/(2\pi)\) must be converted
before identifying their oscillator coefficients with
\(a_\lambda(\vec k)\).
For example, the frequency-angle expansion with prefactor
\(\sqrt{4\pi G/\omega}\,d\omega/(2\pi)\) has mode coefficient
\(\sqrt{2\pi\omega}\,a_\lambda(\omega\widehat{\mathbf k})\).
Its commutator is \(2\pi\delta(\omega-\omega')\delta_{S^2}\).
The Jacobian and Fourier factor reproduce
Eq.~\eqref{eq:NP-shear-momentum-normalized} in the momentum CCR
convention.

For positive-frequency test functions supported away from zero,
normal ordering and the time integral give
\begin{equation}
 \int_{-\infty}^{\infty}du\,:
 \dot\sigma\dot{\bar\sigma}:
 =4\pi G\sum_{\lambda=\pm}
   \int_0^\infty d\omega\,\omega^2
   a_\lambda^\dagger(\omega,\Omega)a_\lambda(\omega,\Omega).
 \label{eq:NP-News-flux-oscillator-check}
\end{equation}
The terms with two annihilators or two creators have support at
\(\omega+\omega'=0\) and are absent on this domain. Substitution
in Eq.~\eqref{eq:classical-News-four-flux-new}, using
\(d^3k=\omega^2d\omega\,d\Omega\), yields
\(P^a=2\sum_\lambda\int d\mu(k)\,k^a a_\lambda^\dagger a_\lambda\).
It also gives \(P^a|k,\lambda\rangle=k^a|k,\lambda\rangle\).
All dynamical kernels below are expressed in this same momentum
CCR convention.
The external Fock states are
\begin{equation}
 |k,\lambda\rangle=a_\lambda^\dagger(k)|0\rangle,
 \qquad
 |X_n\rangle=\prod_{j=1}^n a_{\lambda_j}^\dagger(k_j)|0\rangle.
 \label{eq:QS-I-external-state-normalization}
\end{equation}
Their one-particle overlap is
$\omega_k\delta_{\lambda\lambda'}\delta^{(3)}(\vec k-\vec k')$.
The quantum counterpart of
Eq.~\eqref{eq:DKL-classical-flux-equality-new} is the additional
balance condition introduced below.

\subsection{Observable transport and the classical balance condition}
\label{subsec:quantum-logical-sequence}
\label{subsec:classical-transport-balance}

Let $\mathscr C:\Gamma_-\to\Gamma_+$ be the classical NSF map of
Sec.~\ref{subsec:classical-NSF-flux-condition}, with the asymptotic
identifications already included. Its pullback defines the outgoing
energy as a function of incoming data,
\begin{equation}
 (\mathscr C^*\mathcal H_+)(x_-)
 :=\mathcal H_+\bigl(\mathscr C(x_-)\bigr).
 \label{eq:classical-energy-pullback}
\end{equation}
This definition does not imply $\mathscr C^*\mathcal H_+=\mathcal H_-$:
the latter is the balance requirement
\eqref{eq:NSF-classical-dynamical-energy-condition}, whose direct NSF
verification is formulated in Eq.~\eqref{eq:NSF-classical-flux-exact-check}.
With $H_{\rm in}=P_-^0$ and $H_{\rm out}=P_+^0$, the corresponding
quantum relations in the common representation are
\begin{equation}
 \begin{aligned}
 \mathcal H_+\circ\mathscr C=\mathscr C^*\mathcal H_+
 &\quad\longleftrightarrow\quad H_{\rm out}=S^\dagger H_{\rm in}S,
 \\
 \mathscr C^*\mathcal H_+=\mathcal H_-
 &\quad\longleftrightarrow\quad H_{\rm out}=H_{\rm in}.
 \end{aligned}
 \label{eq:classical-quantum-transport-balance-analogy}
\end{equation}
The first row transports the observable; the second imposes energy
balance. We formulate both as matrix-element relations on the common
domain below, and test the NSF mean energy without imposing the second
in Sec.~\ref{sec:perturbative-scalar-condition}.

\subsection{The two four-momentum relations and their consequence}
\label{subsec:ward-and-dkl}

Under the common-Fock unitary-implementation assumption, the NSF
input--output map is implemented by
$S$ in our convention. For corresponding particle states, we also
assume that the vacuum is invariant under this implementation:
\begin{equation}
 a_{\rm out}=S^\dagger a_{\rm in}S,
 \qquad |\Psi_{\rm out}\rangle=S^\dagger|\Psi_{\rm in}\rangle.
 \label{eq:aout-common-fock-ward}
\end{equation}
Inserting this map into Eq.~\eqref{eq:LSZ-H-check} gives the first
four-momentum scalar-product relation, for arbitrary, independently
chosen $\Phi,\Psi\in\mathcal D_{\rm rad}$,
\begin{equation}
 \langle\Phi|(P_+^a-S^\dagger P_-^aS)|\Psi\rangle=0.
 \label{eq:charge-common-fock}
\end{equation}
The corresponding intertwining relation is understood in the same
scalar-product sense on the common invariant domain.

The DKL matching relation for $\chi$, through its time- and
space-translation components and quantum correspondence, yields
equality of the incoming and outgoing quantum four-momentum flux
operators as a condition on the NSF scattering map. The quantum
counterpart of Eq.~\eqref{eq:DKL-classical-flux-equality-new} is therefore
imposed separately from unitary transport, in the form
\begin{equation}
 \langle\Phi|(P_+^a-P_-^a)|\Psi\rangle=0.
 \label{eq:DKL-quantum-fourmomentum-equality}
\end{equation}
The vanishing soft translation projection was established in
Eq.~\eqref{eq:l-harmonic-property-new}; it does not imply preservation
of an infrared sector. In particular, fixed-$\chi$ classical leaves
and fixed-soft-charge Fock representations are distinct constructions
\cite{AshtekarCampigliaLaddha2018}. Our operator statements concern only
the common radiative representation. The density argument following
Eq.~\eqref{eq:weak-to-operator} gives equality on its common domain;
equality of closures additionally requires a common core.

The temporal components define the Hamiltonians:
\begin{equation}
 H_-\equiv P_-^0,\qquad H_+\equiv P_+^0.
 \label{eq:Hpm-definition}
\end{equation}
Since $l^0=1$, Eqs.~\eqref{eq:charge-common-fock} and
\eqref{eq:DKL-quantum-fourmomentum-equality} yield, respectively,
\begin{equation}
 \langle\Phi|(H_+-S^\dagger H_-S)|\Psi\rangle=0,
 \label{eq:Hplus-definition}
\end{equation}
\begin{equation}
 \langle\Phi|(H_+-H_-)|\Psi\rangle=0.
 \label{eq:DKL-quantum-H-equality}
\end{equation}
The common domain is built from normalizable wave packets; sharp
energy eigenstates are used in the distributional sense.

Subtracting the two relations gives, on this domain,
\begin{equation}
 \langle\Phi|(S^\dagger P_-^aS-P_-^a)|\Psi\rangle=0.
 \label{eq:NSF-translation-condition}
\end{equation}
Its temporal component is the corresponding Hamiltonian condition.
If $S$ and $S^\dagger$ preserve a suitable common invariant domain,
unitarity then gives
$\langle\Phi|[P_-^a,S]|\Psi\rangle=0$ and
$\langle\Phi|[H_-,S]|\Psi\rangle=0$.  Operator commutation
on that domain follows by the same density argument.

In particular, let $|f\rangle_{\rm out}=S^\dagger|f\rangle$, so
${}_{\rm out}\langle f|=\langle f|S$, with $|f\rangle$ and $|i\rangle$
in the incoming reference representation.  On a common domain preserved
by $S$ and $S^\dagger$, choose the test vectors in the unitary
scalar-product condition to obtain
$\langle f|SH_{\rm out}|i\rangle=\langle f|H_{\rm in}S|i\rangle$.
The additional balance condition then yields
\begin{equation}
 \begin{aligned}
 0&={}_{\rm out}\langle f|(H_{\rm out}-H_{\rm in})|i\rangle\\
  &=\langle f|S(H_{\rm out}-H_{\rm in})|i\rangle\\
  &=\langle f|(H_{\rm in}S-SH_{\rm in})|i\rangle
   =\langle f|[H_{\rm in},S]|i\rangle.
 \end{aligned}
 \label{eq:transported-out-balance-commutator}
\end{equation}
This relation holds for arbitrary test vectors; its specialization to
energy eigenstates is given below.

\subsection{Exclusive four-momentum selection}
\label{subsec:operator-condition}

For generalized four-momentum eigenstates in the incoming reference
representation, write
\begin{equation}
 P_-^a|i\rangle=p_i^a|i\rangle,
 \qquad\langle f|P_-^a=p_f^a\langle f|,
 \qquad S_{fi}\equiv\langle f|S|i\rangle.
 \label{eq:common-fourmomentum-eigenstates}
\end{equation}
For delta-normalized states the following equations have their usual
wave-packet or distributional interpretation.  Taking
$\langle\Phi|=\langle f|S$ and $|\Psi\rangle=|i\rangle$
in Eq.~\eqref{eq:NSF-translation-condition} on the invariant domain
gives $\langle f|[P_-^a,S]|i\rangle=0$, and hence
\begin{equation}
 (p_f^a-p_i^a)S_{fi}=0.
 \label{eq:NSF-fourmomentum-selection}
\end{equation}
Thus each nonzero exclusive amplitude has support at $p_f^a=p_i^a$.
In particular, for $H_-|i\rangle=E_i|i\rangle$ and
$\langle f|H_-=E_f\langle f|$,
\begin{equation}
 (E_f-E_i)S_{fi}=0.
 \label{eq:exclusive-energy-conservation}
\end{equation}
The label $E_f$ is the energy of one exclusive final configuration; it
is distinct from the inclusive outgoing expectation value defined below.
For the explicit NSF kernels, the spatial momentum delta functions
supply the spatial part of this support, while the energy selection
condition supplies its temporal part.

For the transported bra
${}_{\rm out}\langle f|=\langle f|S$:
unitary conjugation alone gives
${}_{\rm out}\langle f|(H_+-S^\dagger H_-S)|i\rangle
=E_fS_{fi}-E_fS_{fi}=0$.
The nontrivial selection rule follows from the additional balance
condition, as derived in Eq.~\eqref{eq:transported-out-balance-commutator}.

\subsection{Remark on comparison with Strominger's quantum relations}
\label{subsec:ward-strominger-translations}

Both Strominger's approach and the present construction postulate a
quantum extension of a classical matching result. In
Ref.~\cite{Strominger2014}, the classical matching of the past and future
supertranslation charges motivates their quantum scattering relation.
Strominger explicitly assumes in Sec.~3.3 that the classical and
semiclassical relations extend to scattering in a Fock basis. His
semiclassical argument in Sec.~3.1 also invokes the abelian charge
algebra and the construction of the scattering operator from
exponentials of the Hamiltonian.

Our assumptions are that the radiative fields at past and future null
infinity admit a common Fock representation, that their asymptotic
operators are related by a unitary scattering operator $S$, and that the
classical flux balance implied by mass-aspect continuity has the quantum
counterpart $H_+=H_-$, in the sense specified above. The last statement
is a quantum postulate motivated by the classical result. Combining it
with $H_+=S^\dagger H_-S$ yields $[H_-,S]=0$.
The operator $S$ also implements the scattering dynamics of the NSF
construction: its perturbative generators are reconstructed from the
dynamical map relating incoming and outgoing radiative operators.
The resulting amplitudes therefore provide a dynamical test of energy
balance, evaluated below in the retained channels without imposing an
external energy-conserving delta function.

For comparison, denote Strominger's scattering map by $\mathcal S$.
With compatible identifications of the asymptotic charges and
$\mathcal S=S^\dagger$ to account for our adjoint convention, the
time-translation component of his Eq.~(3.11) reads
\begin{equation}
 H_+\mathcal S=\mathcal S H_-
 \quad\Longleftrightarrow\quad
 H_+=\mathcal S H_-\mathcal S^\dagger
     =S^\dagger H_-S.
 \label{eq:strominger-ward-standard}
\end{equation}
This relation is satisfied identically by unitary transport in our
common Fock representation. Its automatic character here reflects our
assumptions and operator identifications. The quantum balance postulate
$H_+=H_-$ supplies the additional condition that implies commutation
with $S$. This comparison concerns only the time-translation component
of the Ward identity.

\section{Physical amplitudes and the BMS translation subgroup}
\label{sec:physical-amplitudes}

The preceding argument identifies the energy support required if
quantum flux balance is imposed. Since this support is not explicit
in the NSF kernels, we first implement it as an external amplitude
prescription. This defines the constrained calculation and supplies
a reference for the independent dynamical test in the next section.

We restrict attention to the four-dimensional translation subgroup of
BMS. The prescription below enforces conservation of its four charges;
it does not select a Poincar\'e subgroup, impose Lorentz-charge
conservation, or impose the full set of BMS supertranslation constraints.
The three spatial components already appear explicitly in the NSF
kernels, so the additional restriction concerns only total external energy.

Factor the total spatial-momentum distribution from the connected NSF
amplitude and denote its reduced coefficient by $\mathcal M^{\rm NSF}_{fi}$.
The physical prescription considered here is
\begin{equation}
 \boxed{\mathcal A^{\rm phys}_{fi}
 =\delta^{(3)}(\mathbf P_f-\mathbf P_i)\,
 \delta(E_f-E_i)\,
 \left.\mathcal M^{\rm NSF}_{fi}\right|_{E_f=E_i}.}
 \label{eq:physical-amplitude-prescription}
\end{equation}
Overall Fourier and state-normalization factors are absorbed into
$\mathcal M^{\rm NSF}_{fi}$. The prescription acts on the total external
energies after evaluating the NSF kernels and contractions. It introduces
no additional energy delta at an internal vertex. All external gravitons
remain null, $k_j^0=|\mathbf k_j|$. This is the external four-momentum
restriction used in the benchmark calculations of
Refs.~\cite{KDI2026,KDII2026,KDIII2026}; there too, this restriction
concerns only the four translation charges. The Poincar\'e reduction
used for the Weinberg benchmark in Part~I additionally provides the
Lorentz covariance that makes the center-of-momentum calculation
representative of the corresponding invariant kinematics. Conservation
of the BMS translation charges alone does not establish that property.

The energy support in Eq.~\eqref{eq:physical-amplitude-prescription}
is imposed. This prescription neither proves that the NSF unitary
implementer commutes with the incoming Hamiltonian nor guarantees
that imposing the energy delta preserves unitarity.

For a complete tree-level $2\to2$ calculation, the BCH organization is
\begin{equation}
 \mathcal A^{\rm NSF}_{2\to2,\,\mathrm{tree}}
 =\left\langle f\left|
 i\delta T_2-\frac12(\delta T_1)^2
 \right|i\right\rangle_{\mathrm{connected,\ tree}}.
 \label{eq:complete-tree-amplitude}
\end{equation}
Two cubic insertions supply the exchange diagrams and the quartic part
of $\delta T_2$ supplies the contact diagram. Both contributions have
the same perturbative order. The parameter $\varepsilon$ only records
that order and is set to one; it supplies no numerical suppression of
the contact term. Comparison with the complete expression requires specifying the
benchmark truncations and retained kernels.

\begin{figure}[t]
\centering
 
 \includegraphics[width=0.85\linewidth]{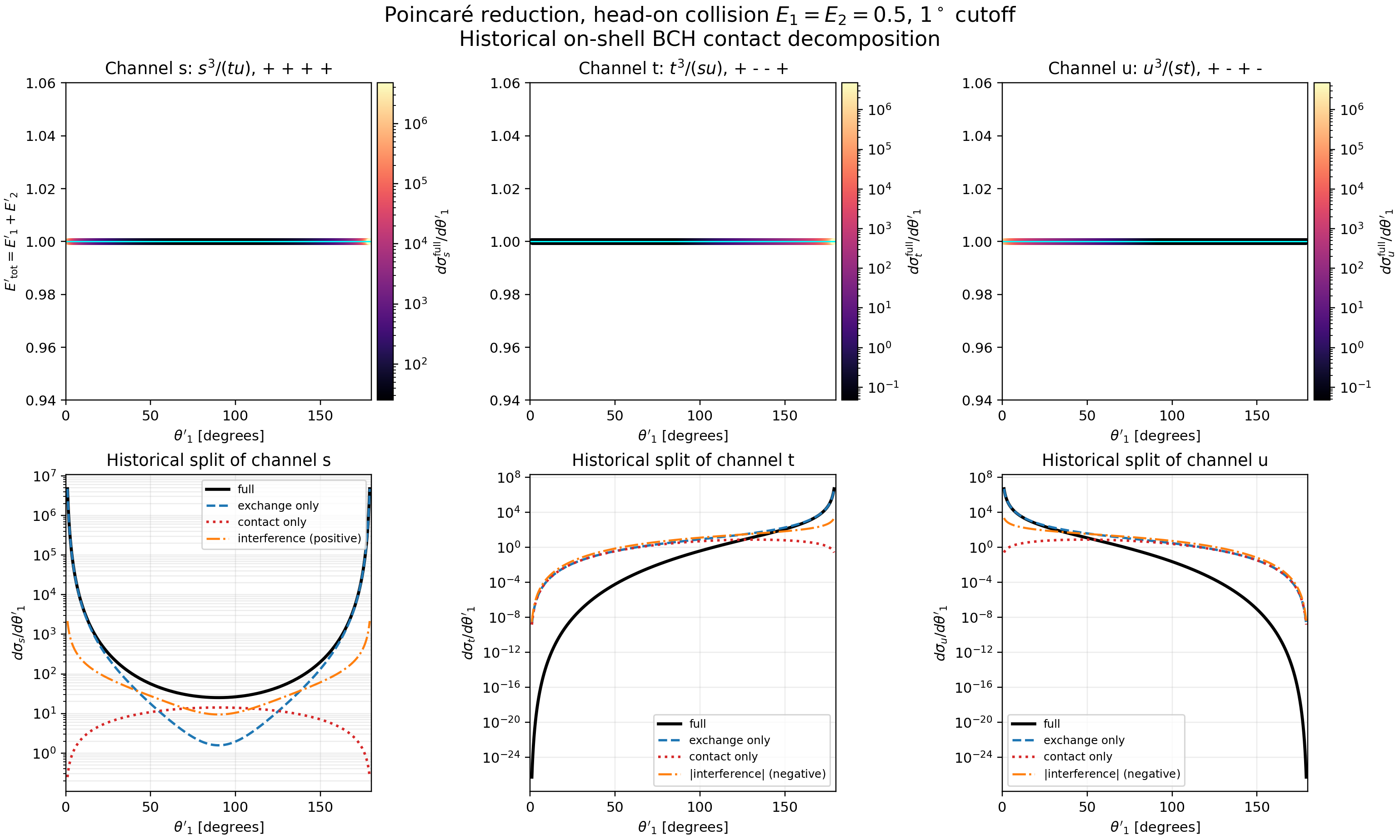}
\caption{Numerical scattering amplitude with the BMS translation-sector constraint of Eq.~\eqref{eq:physical-amplitude-prescription}. Total external energy
 conservation is part of the definition used for this calculation.
 This figure illustrates the physical amplitude, rather than testing
 energy conservation independently.}
 \label{fig:translation-amplitude}
\end{figure}

\section{Numerical mean-energy and variance test without imposing energy conservation}
\label{sec:perturbative-scalar-condition}

The preceding section defined a physical amplitude by imposing total
external energy conservation on the NSF result. The NSF amplitudes
already enforce total linear-momentum conservation through their
dynamical kernels. We now investigate whether energy conservation is
likewise encoded in the scattering dynamics, although not manifest as
an external energy delta. For this purpose, we calculate the conditional
mean outgoing energy and its variance using $S_1=\exp(i\delta T_1)$,
without imposing the additional energy constraint. We first give the constrained mean
as a reference, then specify the $S_1$ calculation and its numerical
implementation.

\subsection{Mean energy when exclusive conservation is imposed}

For an incoming state of definite total energy $E_i$, the prescription
in Eq.~\eqref{eq:physical-amplitude-prescription} restricts every retained
final configuration to $E_f=E_i$. With the channel weights and
energy-weighted integrals defined below, this gives directly
\begin{equation}
 W_n^{\rm phys}=E_iP_n^{\rm phys},\qquad
 \overline E_{\rm out}^{\{2,3\},\rm phys}
 =\frac{W_2^{\rm phys}+W_3^{\rm phys}}
 {P_2^{\rm phys}+P_3^{\rm phys}}=E_i.
 \label{eq:physical-two-three-energy-exact}
\end{equation}
The denominator is assumed nonzero, and probabilities are formed with
a common scattering regulator before taking the sharp-energy limit.
This equality follows from imposing exclusive energy conservation;
it is not an independent prediction of the NSF amplitudes. A selection
of outgoing channels can reweight the energy components of an incoming
packet, so the statement here concerns fixed incoming energy.

\subsection{Removing the exclusive energy condition}

We now omit the additional equality $H_{\rm out}=H_{\rm in}$ and do not
apply Eq.~\eqref{eq:physical-amplitude-prescription}. The spatial-momentum
distributions and radiative on-shell conditions already present in NSF
are retained. The question is whether the unmodified amplitudes yield
an outgoing mean energy close to the incoming energy.

\subsection{Mean outgoing energy using $S_1=\exp(i\delta T_1)$}

We calculate the mean outgoing energy using the unitary operator
\begin{equation}
 S_1=\exp(i\delta T_1),\qquad \delta T_1^\dagger=\delta T_1.
 \label{eq:numerical-S1-model}
\end{equation}
With the assumed self-adjoint realization of $\delta T_1$, this operator
implements the outgoing modes and Hamiltonian as
\begin{equation}
 a_{{\rm out},1}=S_1^\dagger a_{\rm in}S_1,\qquad
 H_{{\rm out},1}=S_1^\dagger H_{\rm in}S_1.
 \label{eq:numerical-S1-observables}
\end{equation}
For a normalized incoming state the full mean is
$\langle i|H_{{\rm out},1}|i\rangle$. The numerical calculation
reported here evaluates its leading connected $2\to2$ and $2\to3$
channel amplitudes,
\begin{align}
 \mathcal A_{22}^{\rm lead}
 &=-\frac12\langle f_2|(\delta T_1)^2|i_2\rangle_{\rm conn},\\
 \mathcal A_{23}^{\rm iter}
 &=-\frac{i}{6}\langle f_3|(\delta T_1)^3|i_2\rangle_{\rm conn}.
 \label{eq:numerical-retained23}
\end{align}
Their squared moduli, averaged over incoming helicities and summed
over outgoing helicities, determine the channel weights and energy
moments defined below. All retained routings and intermediate
helicities are combined coherently before squaring. No additional
external energy delta is imposed, and the exact spatial zero-mode
fusion routing is excluded in the CoM radiative domain.

The use of $S_1$ is motivated in part by the $2\to2$ comparison in
Fig.~\ref{fig:translation-amplitude}. The integrated angular probability
is dominated by the forward and backward peaks, where the full and
exchange-only curves nearly coincide. The independent $\delta T_2$
contact contribution and its interference have a small effect on that
integrated benchmark, even though they can be important locally at
angles with much smaller probability. This motivates the approximation
adopted here. The present energy test uses $S_1$ and the specified
angular-kernel prescription throughout.

The reported observable is the conditional mean in these two leading
connected channels. The exact exponential $S_1$ also contains higher
powers of $\delta T_1$ and other final sectors, so its unitarity does
not normalize the sum of the two truncated weights or establish energy
balance. Appendix~\ref{app:connected-channel-energies} supplies the
normalization and energy integrals; Appendix~\ref{app:numerical-implementation}
gives the numerical prescription and checks. The modifications produced
by independent $\delta T_2$ and $\delta T_3$ generators are discussed
separately in Appendix~\ref{app:higher-generator-extension}.

\subsection{Conditional mean outgoing energy}
\label{subsec:conditional-mean-definition}

For each incoming helicity configuration, let $|i\rangle$ be a
normalized two-graviton state with fixed total energy
$E_{\rm in}=\langle i|H_{\rm in}|i\rangle=E_i$. Sharp momentum states
are understood through a common wave-packet or finite-volume regulator
before probabilities are formed. The unpolarized calculation averages
the squared amplitude over the four incoming helicity configurations
and sums over all unobserved outgoing helicities:
\begin{equation}
 \overline{|\mathcal A^{\rm NSF}_{2\to n}|^2}
 =\frac14\sum_{\lambda_1,\lambda_2=\pm2}
 \sum_{\lambda'_1,\ldots,\lambda'_n=\pm2}
 \left|\mathcal A^{\rm NSF}_{\lambda_1\lambda_2
 \to\lambda'_1\cdots\lambda'_n}\right|^2,
 \qquad n=2,3.
 \label{eq:numerical-helicity-average-sum}
\end{equation}
The factor $1/4$ applies only to the incoming helicities.
For the retained amplitudes, define the channel weights and energy
moments by
\begin{align}
 P_n&=\int d\Pi_n\,\overline{|\mathcal A^{\rm NSF}_{2\to n}|^2},
 \label{eq:numerical-channel-weight}\\
 W_n&=\int d\Pi_n\,E_n\,\overline{|\mathcal A^{\rm NSF}_{2\to n}|^2},
 \qquad E_n=\sum_{j=1}^{n}|\mathbf k'_j|.
 \label{eq:numerical-channel-energy}
\end{align}
Here $d\Pi_n$ contains the identical-particle factors and the chosen
integration domain in the stated oscillator convention. The helicity
average and sums are included only in the overlined squared amplitude.
No delta of total incoming minus outgoing energy is imposed. The same
regulator, normalization, and integration prescription are used in
$P_n$ and $W_n$.

For $P_2+P_3>0$, the two channel contributions use a common denominator:
\begin{equation}
 \mathcal E_{22}=\frac{W_2}{P_2+P_3},\qquad
 \mathcal E_{23}=\frac{W_3}{P_2+P_3}.
 \label{eq:numerical-channel-energy-contributions}
\end{equation}
Their sum is the energy observable tested numerically,
\begin{equation}
 \boxed{\overline E_{\rm out}^{\{2,3\}}
 =\mathcal E_{22}+\mathcal E_{23}
 =\frac{W_2+W_3}{P_2+P_3},\qquad
 r_E=\frac{\overline E_{\rm out}^{\{2,3\}}-E_{\rm in}}{E_{\rm in}}.}
 \label{eq:numerical-conditional-residual}
\end{equation}
If each retained process conserves total energy, then $W_n=E_iP_n$
and this sum equals $E_i$. In the numerical test, the energy moments
are instead evaluated directly without imposing that condition.

This is the mean conditioned on the two retained connected channels
and the specified acceptance. The full outgoing expectation in the
chosen model is $\langle i|S_1^\dagger H_{\rm in}S_1|i\rangle$
for a fixed incoming helicity state, followed by the stated incoming
helicity average. It includes all sectors and all powers of
$\delta T_1$, unlike the present two-channel leading calculation. At finite diagrammatic truncation,
$P_n$ are the corresponding channel weights; their sum need not equal
one. Appendix~\ref{app:connected-channel-energies} gives the explicit
connected amplitudes, normalization factors, and energy-weighted
integrals for both processes.

\subsection{Numerical preparation and integration prescription}
\label{subsec:numerical-preparation}

The incident momenta in the numerical calculation are exactly
$\mathbf k_1=(E_i/2)\hat{\mathbf z}$ and
$\mathbf k_2=-(E_i/2)\hat{\mathbf z}$, with $E_i=1$ and
$c=\hbar=G=1$. The initial-helicity average and final-helicity sum are
those of Eq.~\eqref{eq:numerical-helicity-average-sum}.

There is no fixed minimum angle on an external momentum pair in the
present calculation. The $1^\circ$ threshold belongs to an earlier
monochromatic $2\to2$ calculation. Here the infrared cutoff restricts
outgoing momentum norms and specified internal spatial-momentum
routings. The sphere integral in a cubic kernel uses the circular
finite-part prescription detailed in
Appendix~\ref{app:numerical-implementation}. Its equivalence to the
heat-regulated or full distributional NSF kernel has not been established.

The spatial delta is resolved exactly. For $2\to2$ the remaining
variables can be taken as $x=E_f/E_i$ and the smaller transfer norm $q$;
for $2\to3$ there are two radial and three angular variables.
In this numerical prescription the upper bound is on the total
outgoing energy, $E_f\le E_{\max}$, rather than on each independent
radial integration separately. The reference infrared scan uses
$E_{\max}/E_i=2024$. Explicit measures, routing cuts, and the reciprocal
$2\to2$ coordinate transformation are given in
Appendix~\ref{app:numerical-implementation}.

The direct low-cutoff runs use eight independent scrambled Sobol
replicas per energy stratum for $2\to3$ and deterministic
Gauss--Legendre quadrature for $2\to2$. Refinement concerns the
geometry-dependent quadrature panels and selected order-doubling
checks; sample sizes are prescribed in advance. There is no stopping
rule that increases samples until a target error is reached. The
$10^{-5}$ result uses a separately documented four-replica archived
profile. Appendix~\ref{app:numerical-implementation} records both
profiles and the covariance propagation for the energy residual.

\subsection{Mean-energy residual and outgoing-energy variance}
\label{subsec:numerical-energy-and-variance}

The two principal observables are the conditional mean-energy residual
$r_E$ of Eq.~\eqref{eq:numerical-conditional-residual} and the physical
variance of $X=E_f/E_i$. In the same retained channels and with the same
integration prescriptions, let
\begin{equation}
 C_n=\int d\Pi_n\,(X-1)^2\,
 \overline{|\mathcal A^{\rm NSF}_{2\to n}|^2},\qquad
 v\equiv\operatorname{Var}(X)
 =\frac{C_2+C_3}{P_2+P_3}-r_E^2.
 \label{eq:numerical-variance-highlight}
\end{equation}
The corresponding complete channel-moment definitions, covariance
propagation, and checks appear in Appendix~\ref{app:energy-variance}.
The variance measures the physical energy spread, not the Monte Carlo
standard error of its estimator.

At $\kappa_{\rm IR}/E_i=10^{-8}$ and fixed $E_{\max}=2024E_i$,
\begin{equation}
 \boxed{\frac{\overline E_{\rm out}^{\{2,3\}}}{E_i}
 =1.000000611\pm0.000000047,\quad
 r_E=(6.11\pm0.47)\times10^{-7},\quad
 v=(4.020\pm0.337)\times10^{-5}.}
 \label{eq:numerical-energy-result}
\end{equation}
The reported uncertainties are statistical standard errors, while $v$
is the physical variance. At fixed $E_{\max}$, lowering the infrared
cutoff from $10^{-5}$ to $10^{-8}$ reduces the central residual by a
factor of about 584 and the variance from
$(3.040\pm0.784)\times10^{-2}$ to
$(4.020\pm0.337)\times10^{-5}$, a factor of about 756.
Thus the conditional distribution becomes narrower around $E_i$ in
this finite-cutoff numerical scan, rather than merely reproducing the
incoming energy in its first moment. These data do not establish a
zero-cutoff limit or exclusive energy conservation.

\begin{figure}[t]
\centering
\includegraphics[width=0.85\linewidth]{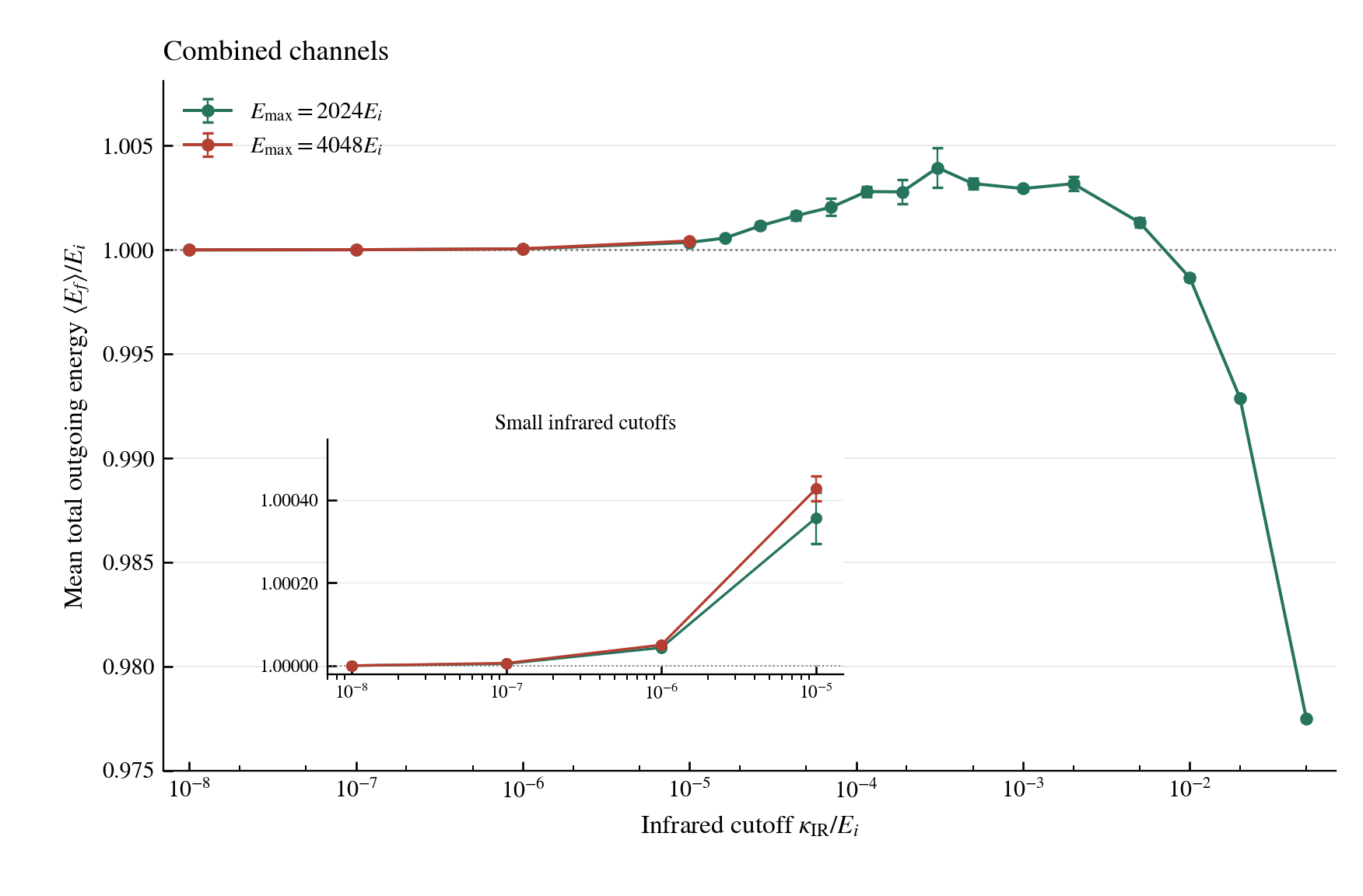}
\caption{Conditional mean total outgoing energy,
$\langle E_f\rangle/E_i$, in the retained $2\to2$ and $2\to3$
channels, combined according to their calculated relative weights,
as a function of the infrared cutoff $\kappa_{\mathrm{IR}}/E_i$.
The green and red curves correspond to $E_{\max}=2024E_i$ and
$E_{\max}=4048E_i$, respectively. The inset enlarges the region
of small infrared cutoffs. The dotted line marks
$\langle E_f\rangle=E_i$; no external energy-conserving delta
function is imposed. Calculations use $E_i=G=\epsilon=1$
with the retained $2\to2$ exchange and iterated-cubic $2\to3$
amplitudes of the cubic-generator model $S_1=\exp(i\delta T_1)$.
Error bars represent one statistical standard error of the
estimated mean and exclude uncertainties associated with the
amplitude truncation and angular prescription. Lines connecting
the evaluated points guide the eye. All plotted points use the four
infrared cutoffs $\kappa_{\mathrm{IR}}/E_i=10^{-5},10^{-6},10^{-7},10^{-8}$
and the reference infrared-scan results described in the text.
The $E_{\max}=4048E_i$ curve shows the earlier local estimate at
$\kappa_{\mathrm{IR}}/E_i=10^{-8}$; the updated paired estimate is
reported in Table~\ref{tab:nsf-upper-energy-bound}.}
\label{fig:energia_media_combinada}
\end{figure}

The full cutoff scan and the channel-resolved diagnostics, including
$r_2$, $r_3$ and their contributions to $r_E$, are documented in
Appendix~\ref{app:numerical-implementation}; see
Table~\ref{tab:nsf-ir-channel-weights} and
Fig.~\ref{fig:unreduced-energy-residual}. The separate $2\to3$
weight rises from $94.13\%$ at $\kappa_{\rm IR}/E_i=10^{-5}$ to
$99.99330\%$ at $10^{-8}$. Hence the combined residual must be
interpreted together with the changing channel weights. The
channel-resolved data are retained as checks against cancellations
between channels, but are not independent conservation laws.

\subsection{Infrared cutoff and upper-energy sensitivity}
\label{subsec:numerical-infrared-cutoff}
\label{subsec:numerical-upper-energy-bound}

The reference scan fixes $E_{\max}=2024E_i$, with the upper
bound applied to the total outgoing energy. The infrared and angular
prescriptions are specified in Sec.~\ref{subsec:numerical-preparation}.

At $\kappa_{\rm IR}/E_i=10^{-8}$, increasing
$E_{\max}/E_i$ from 2024 to 18096 changes $r_E$ from
$(6.114\pm0.466)\times10^{-7}$ to
$(6.612\pm0.494)\times10^{-7}$. Over the same interval the
variance increases from $4.020\times10^{-5}$ to
$4.106\times10^{-4}$, a factor of approximately 10.21. The second
moment is therefore substantially more sensitive to the upper-energy
bound than the first. Complete paired comparisons and systematic
checks are retained in Table~\ref{tab:nsf-upper-energy-bound} and
Appendix~\ref{app:energy-variance}. These finite-bound calculations do
not establish an infinite-upper-bound or joint infrared--ultraviolet
limit. Neither statistical errors nor cutoff variation quantify
uncertainties from omitted higher-generator amplitudes.

\section{Discussion and conclusions}
\label{sec:discussion-outlook}
\label{sec:conclusions}

In this work, we have formulated the quantum counterpart of the classical
Bondi four-momentum flux-balance law within the Null Surface Formulation
(NSF) of general relativity. The classical relation, established in the
work of Dom\'inguez, Kozameh, and Ludvigsen
\cite{DominguezKozamehLudvigsen1996,DominguezKozamehLudvigsen1997}
under the appropriate matching and endpoint conditions, provides four
conservation conditions relating incoming and outgoing radiative
fluxes. Their quantum formulation allows an independent test of the
NSF scattering dynamics, since these conditions were not imposed when
the dynamical scattering kernels were constructed.

A principal result is that three of the four conservation conditions
are already satisfied by the quantum NSF scattering amplitudes. These
correspond to conservation of the three components of total spatial
momentum, which is explicit in the amplitudes without an additional
constraint. This agreement between the independently formulated
quantum flux-balance conditions and the NSF dynamics provides a
nontrivial consistency check in the scattering sectors considered;
it does not establish the consistency of the full quantum theory.
The remaining condition, conservation of total energy, is not
manifest in the dynamical kernels and remains an open analytical
problem.

We have considered two distinct approaches to this condition. The
first is to define physical scattering amplitudes by implementing
the energy-conservation requirement derived from quantum flux balance.
The second is to determine whether the same requirement follows
from the unmodified NSF dynamical equations, without introducing
an additional constraint. Establishing the latter would complete
the four-momentum consistency test directly at the dynamical level.

To examine this second possibility, we have performed a limited
numerical test using the unitary approximation generated by
$S_1=\exp(i\delta T_1)$, retaining the connected $2\to2$ and
$2\to3$ channels. At fixed upper-energy cutoff, both the relative
residual of the conditional mean outgoing energy and the variance of
the conditional outgoing-energy distribution decrease substantially
as the infrared cutoff is reduced. These observations are compatible
with the expected energy-balance condition, but they do not establish
a zero-cutoff limit or energy conservation for the complete NSF
scattering amplitudes. In particular, the omitted higher-order
contributions and the dependence on the integration prescriptions
limit the inference. The numerical test is intended to motivate an
analytical investigation, rather than to provide a proof.

One possible direction is to reconsider the stage at which the
four-dimensional NSF dynamics is projected onto radiative modes. In
the LSZ-type organization adopted here, this projection is applied to
scattering matrix elements assembled from the NSF mode variations,
rather than separately to the individual variations
$\delta a_2,\delta a_3,\ldots$ before forming each amplitude. It would
be useful to investigate whether the four-dimensional translation
subgroup of BMS permits the dependence on total incoming and outgoing
momenta to be factored out before the final radiative projection.
Such a factorization is not guaranteed by this organization and would
have to be established from the NSF equations; nevertheless, it offers
a concrete route toward the fourth conservation condition.

An independent verification of the classical Bondi flux-balance law
in the nonlinear classical graviton-scattering construction of
Ref.~\cite{BordcochKozamehRojas2023} would also be worthwhile. In
that formulation, spatial-momentum conservation is not exhibited
through explicit momentum-conserving delta distributions, whereas
the operator commutation relations and momentum-state matrix elements
make the corresponding structure explicit in the quantum amplitudes.
This difference concerns the representation of the calculation,
not the validity of the classical balance law.

Thus, the main result is the formulation of quantum Bondi flux balance
and the independent agreement of three of its four conservation
conditions with NSF scattering dynamics. The analytical origin of
energy conservation in the unmodified NSF amplitudes remains the
central question for future work.

\section*{Data availability}
The numerical code, computational documentation, and reference summary
for the numerical analysis are publicly available at
\url{https://github.com/GerDep/nsf-energy}.
The repository provides scripts for the retained $2\to2$ and $2\to3$
channels, the conditional outgoing-energy mean and variance, and
numerical consistency checks. The availability of the underlying
weighted samples is described in the repository documentation.

\appendix
\section{Connected amplitudes and channel contributions to the mean energy}
\label{app:connected-channel-energies}

\subsection{Leading connected channels of the cubic-generator model}

The model used in the numerical calculation is
$S_1=\exp(i\delta T_1)$. For ordering its exponential one may write
$S_1(\varepsilon)=\exp(i\varepsilon\delta T_1)$ and set
$\varepsilon=1$ afterward. Here $X_N$ denotes the outgoing momenta
and helicities. The leading connected matrix elements are
\begin{align}
 \mathcal A_{22}^{\rm ret}
 &=-\frac{\varepsilon^2}{2}
 \langle X_2|(\delta T_1)^2|i\rangle_{\rm conn},\\
 \mathcal A_{23}^{\rm ret}
 &=-\frac{i\varepsilon^3}{6}
 \langle X_3|(\delta T_1)^3|i\rangle_{\rm conn}.
 \label{eq:appendix-retained-amplitudes}
\end{align}
Each amplitude sums its connected routings and intermediate helicities
before squaring. The full unitary model also contains higher powers
of $\delta T_1$ and additional final sectors; these are distinct from
introducing independent generators $\delta T_2,\delta T_3,\ldots$.

For clarity, write $\mathcal A_{G,2N}$ for the reduced amplitudes in the
oscillator normalization of the quadratures, including the external
frequency factors and with the common spatial normalization removed.
For a helicity-dependent integrand $F(X_N)$, define
\begin{equation}
 \mathcal J_N[F]=\frac{1}{4N!}
 \sum_{\lambda_1,\lambda_2}
 \sum_{\lambda'_1,\ldots,\lambda'_N}
 \int_{\mathcal D_N}\delta^{(3)}\!\left(\sum_a\mathbf p_a\right)
 F(X_N)\prod_a d^3p_a.
 \label{eq:appendix-channel-functional}
\end{equation}
The factor $1/4$ is the incoming-helicity average and $1/N!$ the
identical-final-particle factor. No extra incoming factorial is used.
The spatial constraint is resolved exactly, not squared as an ordinary
function. The relation to the invariant measure $d\mu=d^3p/(2\omega)$
requires the compensating factors $\prod_a2\omega_a$ shown in
Eq.~\eqref{eq:num-measure}; it must not be replaced by an additional
$2^N$ factor on an amplitude already converted to the quadrature
normalization. Appendix~\ref{app:numerical-implementation} specifies
the external factors and reduced variables. The amplitudes in the
following integrals are $\mathcal A_{G,22}$ and $\mathcal A_{G,23}$.
In the common reduced normalization, $P_N=\varepsilon^{2N}I_{2N}$;
all displayed numerical means use $\varepsilon=1$.

The channel weights and their energy moments are
\begin{align}
 P_2&=\mathcal J_2[|\mathcal A_{G,22}|^2],&
 W_2&=\mathcal J_2[(\omega'_1+\omega'_2)|\mathcal A_{G,22}|^2],
 \label{eq:appendix-channel22-moments}\\
 P_3&=\mathcal J_3[|\mathcal A_{G,23}|^2],&
 W_3&=\mathcal J_3[(\omega'_1+\omega'_2+\omega'_3)
 |\mathcal A_{G,23}|^2].
 \label{eq:appendix-channel23-moments}
\end{align}
For $P_2+P_3>0$, the contributions of the two processes to the
conditional mean use the same normalization:
\begin{align}
 \mathcal E_{22}
 &=\frac{\mathcal J_2[(\omega'_1+\omega'_2)|\mathcal A_{G,22}|^2]}
 {P_2+P_3}=\frac{W_2}{P_2+P_3},
 \label{eq:appendix-mean22}\\
 \mathcal E_{23}
 &=\frac{\mathcal J_3[(\omega'_1+\omega'_2+\omega'_3)
 |\mathcal A_{G,23}|^2]}
 {P_2+P_3}=\frac{W_3}{P_2+P_3}.
 \label{eq:appendix-mean23}
\end{align}
The probability weight in each integral is the squared modulus of the
corresponding complete retained amplitude. Adding the two contributions
gives the conditional mean outgoing energy,
\begin{equation}
 \overline E_{\rm out}^{\{2,3\}}
 =\mathcal E_{22}+\mathcal E_{23}
 =\frac{W_2+W_3}{P_2+P_3}.
 \label{eq:appendix-combined-channel-mean}
\end{equation}
If total external energy is conserved in each process at fixed incoming
energy $E_i$, then $W_2=E_iP_2$ and $W_3=E_iP_3$, and hence
\begin{equation}
 \overline E_{\rm out}^{\{2,3\}}
 =\frac{E_iP_2+E_iP_3}{P_2+P_3}=E_i.
 \label{eq:appendix-conserved-channel-sum}
\end{equation}
In the numerical test without an imposed energy delta, both energy
moments are evaluated directly. The separate channel means appearing
in the channel-residual plots are $W_2/P_2$ and $W_3/P_3$ when the
respective weights are nonzero; the contributions
$\mathcal E_{22}$ and $\mathcal E_{23}$ above include the relative
channel weights in the selected ensemble.
These are connected channel weights; identity and spectator
contributions are outside the selected ensemble. Combining the leading
weights of the two channels does not constitute a complete probability
expansion through order $\varepsilon^6$.

\subsection{Extension to independent higher generators}
\label{app:higher-generator-extension}

To extend the cubic-generator model, replace $S_1$ by
\begin{equation}
 S=\exp\!\left[i\left(\varepsilon\delta T_1
 +\varepsilon^2\delta T_2+\varepsilon^3\delta T_3+\cdots\right)\right].
\end{equation}
The parameter $\varepsilon$ records the recursive order and is set to
one. The corresponding formal connected tree coefficients are
\begin{align}
 \mathcal M_{22}^{(2)}(X_2;i)
 &=\left\langle X_2\left|i\delta T_2
 -\frac12(\delta T_1)^2\right|i\right\rangle_{\rm conn},
 \label{eq:appendix-M22}\\
 \mathcal M_{23}^{(3)}(X_3;i)
 &=\left\langle X_3\left|i\delta T_3
 -\frac12(\delta T_1\delta T_2+\delta T_2\delta T_1)
 -\frac{i}{6}(\delta T_1)^3\right|i\right\rangle_{\rm conn}.
 \label{eq:appendix-M23}
\end{align}
Thus the first independent extension adds $i\delta T_2$ to the
$2\to2$ coefficient and the two ordered mixed products together with
$i\delta T_3$ to $2\to3$. These higher-order contributions have
been explicitly identified but deliberately left unevaluated in the
present numerical study: their calculation requires independent NSF
kernels and a substantially more involved analysis of kinematically
allowed connected contractions, helicity sums, and interference terms.
Their omission defines the $S_1$ approximation; the additional
amplitudes are not assumed to vanish.
For the displayed $2\to2$ benchmark,
the integrated contact contribution and interference are small relative
to the dominant exchange probability, motivating their neglect in the
present $S_1$ calculation. The benchmark comparison does not determine
the corresponding correction over the unrestricted outgoing-energy
domain, or the size of the independent $2\to3$ terms. The total amplitude must be squared after
these terms are added, retaining all their interference. Both channel
weights and energy moments must then be recomputed before forming
$(W_2+W_3)/(P_2+P_3)$. These terms need not be small at $\varepsilon=1$, and their
uncertainties are not included in the reported numerical errors.
The enlarged $S$ must also be used consistently in observable
transport.

The additional kernel assembly, ordered contractions, helicity sums,
and interference integrals motivate an analytical investigation of
energy balance directly from the NSF scattering map.

\section{Numerical implementation and documented checks}
\label{app:numerical-implementation}

This appendix specifies the numerical method and its validation checks. Here $x=E_f/E_i$
denotes outgoing energy, while $k=\kappa_{\rm IR}/E_i$ denotes the
infrared cutoff. The reduced squared amplitudes $\mathcal W_{2n}$
below are distinct from the energy moments $W_n$ in the main text.

\subsection{Channel weights and upper-energy scan}

The following tables collect the quantitative results referenced in
Sec.~\ref{subsec:numerical-energy-and-variance}. They retain the
channel-resolved weights and the paired upper-energy scan, while the
physical second moments are tabulated in
Appendix~\ref{app:energy-variance}.

\begin{table}[t]
 \centering
 \caption{Energy residual and relative weight of the inelastic,
 gravitational bremsstrahlung-type $2\to3$ channel near the residual
 maximum and in the subsequent low-cutoff regime. The residual is
 $r_E=\overline E_{\rm out}^{\{2,3\}}/E_i-1$, and
 $p_{23}=I_{23}/(I_{22}+I_{23})$ is reported as a percentage;
 $p_{22}=1-p_{23}$. The calculation uses $E_i=G=\epsilon=1$ and
 $E_{\max}=2024E_i$. Errors are one statistical standard error.
 The complete tabulated results appear in
 Appendix~\ref{app:numerical-implementation}.}
 \label{tab:nsf-ir-channel-weights}
 \begin{tabular}{ccc}
 \toprule
 $\kappa_{\rm IR}/E_i$ & $r_E$ & $p_{23}\;(\%)$ \\
 \midrule
 $10^{-3}$ & $(2.94\pm0.14)\times10^{-3}$ & $11.3\pm1.0$ \\
 $5\times10^{-4}$ & $(3.18\pm0.26)\times10^{-3}$ & $23.0\pm1.4$ \\
 $10^{-5}$ & $(3.57\pm0.62)\times10^{-4}$ & $94.13\pm0.88$ \\
 $10^{-6}$ & $(4.38\pm0.28)\times10^{-5}$ & $99.405\pm0.034$ \\
 $10^{-7}$ & $(5.24\pm0.27)\times10^{-6}$ & $99.9371\pm0.0043$ \\
 $10^{-8}$ & $(6.11\pm0.47)\times10^{-7}$ & $99.99330\pm0.00043$ \\
 \bottomrule
 \end{tabular}
\end{table}

\begin{table}[t]
 \centering
 \caption{Total outgoing-energy upper-bound scan at
 $\kappa_{\rm IR}/E_i=10^{-8}$. The mean is $1+r_E$.
 Errors are one statistical standard error; differences relative to
 $M=2024$ include the covariance of common samples, where
 $M=E_{\max}/E_i$. The $2\to3$ fractions are unchanged at the displayed
 precision.}
 \label{tab:nsf-upper-energy-bound}
 \begin{tabular}{rccc}
 \toprule
 $M$ & $p_{23}\;(\%)$ & $10^7r_E$ & $10^7\Delta r_E$\\
 \midrule
 2024 & $99.99330\pm0.00043$ & $6.114\pm0.466$ & $0$\\
 4048 & $99.99330\pm0.00043$ & $6.229\pm0.472$ & $0.114\pm0.013$\\
 16192 & $99.99330\pm0.00043$ & $6.592\pm0.493$ & $0.478\pm0.042$\\
 18096 & $99.99330\pm0.00043$ & $6.612\pm0.494$ & $0.498\pm0.044$\\
 \bottomrule
 \end{tabular}
\end{table}

\subsection{Numerical setup}
\paragraph{Momentum cut and angular prescription.}
We use $c=\hbar=G=1$ and normalize all momenta by the incoming total
energy $E_i$. Let $x=E_f/E_i$, $k=\kappa_{\mathrm{IR}}/E_i$, and
$M=E_{\max}/E_i$. The two incident momenta are
$\mathbf k_1=\hat{\mathbf z}/2$ and $\mathbf k_2=-\hat{\mathbf z}/2$.
The present calculation does not impose a fixed minimum angle on any
external momentum pair. The $1^\circ$ threshold was used in an earlier
monochromatic $2\to2$ calculation and does not specify the present domain.
The infrared restriction is instead imposed on the spatial momentum norms.
Each outgoing mode satisfies $|\mathbf p_a|\ge k$.
For $2\to2$, both exchange transfers
$|\mathbf p_1-\mathbf k_1|$ and $|\mathbf p_1+\mathbf k_1|$
must be at least $k$. For $2\to3$, the tested internal routings are the
two-leaf sums of the signed list
$(-\mathbf k_1,-\mathbf k_2,\mathbf p_1,\mathbf p_2,\mathbf p_3)$,
excluding the identically zero incoming-pair sum. Thus the tested pairs
are incoming--outgoing differences and outgoing--outgoing sums, not an
angular condition between outgoing directions. The latter sums are equal
in norm to the complementary outgoing momentum in the center of mass.
The exact zero internal mode is omitted. On the elastic $2\to2$ slice,
the transfer restriction does imply a forward angle
$\theta\ge 2\arcsin k$ on the folded hemisphere; away from that slice
it is a condition on both energy and angle.

The sphere integral inside each cubic vertex uses a circular finite-part
prescription for the unfiltered angular kernel. Around the two daughter
directions and the parent extraction direction we form a smooth partition
of unity. The radius around pole $a$ is
$0.35\min_{b\ne a}\angle(\hat{\mathbf n}_a,\hat{\mathbf n}_b)$.
In each region the azimuthal average is taken before the radial integral
in $1-\cos\theta_a$, whose lower endpoint is zero. The complement is also
integrated. Hence the radii specify a partition of the entire sphere.
Exactly collinear vertices are not assigned ordinary point values.
This prescription specifies the calculation; it does not establish
equivalence to the heat-regulated or full distributional NSF kernel.

\paragraph{Measure and energy bounds.}
The one-particle convention is $\mathrm{d}\mu(p)=\mathrm{d}^3p/(2\omega_p)$,
with $\omega_p=|\mathbf p|$ and no $(2\pi)^{-3}$ in the measure.
For clarity, let $\mathcal W_{2n}=\frac14\sum_{\lambda_i,\lambda_f}|A_{G,2n}|^2$
denote the squared reduced amplitude in the oscillator normalization
used by the quadratures, including the external frequency factors.
The two incident helicities are averaged independently and all final
helicities are summed. For $2\to2$ the external oscillator divisor is
$\sqrt{\omega_{k_1}\omega_{k_2}\omega_{p_1}\omega_{p_2}}=x/4$.
For $2\to3$ the canonical projection already contains the frequency
factors of the cubic vertices, with the common cell prefactor removed.
After removal of the common spatial normalization,
the implemented integral and its invariant-measure form are
\begin{align}
 I_{2n}&=\frac{1}{n!}\int_{\mathcal D_n}
 \delta^{(3)}\!\left(\sum_a\mathbf p_a\right)
 \mathcal W_{2n}\prod_a\mathrm{d}^3p_a \nonumber\\
 &=\frac{1}{n!}\int_{\mathcal D_n}
 \delta^{(3)}\!\left(\sum_a\mathbf p_a\right)
 \left[\left(\prod_a 2\omega_a\right)\mathcal W_{2n}\right]
 \prod_a\mathrm{d}\mu(p_a).
 \label{eq:num-measure}
\end{align}
Here $\mathcal D_n$ includes the momentum restrictions above and
$x=\sum_a\omega_a\le M$. There is a factor $1/2!$ or $1/3!$ for
identical outgoing gravitons, and no additional incoming factorial for
the specified two-mode initial state. The spatial delta is resolved
exactly; there is no external $\delta(E_f-E_i)$.
The bound $M$ is on the \emph{total} final energy. The reference
infrared scan uses $M=2024$; Table~\ref{tab:num-emax} varies this bound
at fixed $k=10^{-8}$.

For $2\to2$, $\mathbf p_2=-\mathbf p_1$ and $x=2|\mathbf p_1|$.
Azimuthal symmetry and exchange of the final labels give, after the
$1/2!$ factor, the integration measure $\pi x^2\mathrm{d} x\,\mathrm{d}\mu/4$
on the folded hemisphere $0\le\mu=\cos\theta\le 1$.
With $q=\tfrac12\sqrt{x^2+1-2x\mu}$ the same integral is
\begin{equation}
 I_{22}=\pi\int_{2k}^{M}\mathrm{d} x\,x
 \int_{\max(k,|x-1|/2)}^{\sqrt{x^2+1}/2}\mathrm{d} q\,q\,\mathcal W_{22}(x,q).
 \label{eq:num-22measure}
\end{equation}
The two transfer restrictions reduce to the lower limit shown because
$q$ is the smaller transfer on this hemisphere. The implementation folds
$x=u$ and $x=1/u$, $u\ge1$, using the reciprocity of the $2\to2$ kernel.
Both branches keep their own $2k\le x\le M$ support. For the reciprocal
branch the rescaled momentum cut is $ku$; the folded upper endpoint is
$\max(M,1/(2k))$. This endpoint is a quadrature coordinate bound, not an
increase of the physical $E_{\max}$.

For $2\to3$, write $\mathbf p_1=\mathbf p$, $\mathbf p_2=\mathbf q$,
$\mathbf p_3=-\mathbf p-\mathbf q$. Then
$x=|\mathbf p|+|\mathbf q|+|\mathbf p+\mathbf q|$ and
$I_{23}=\frac{1}{3!}\int_{\mathcal D_3}\mathrm{d}^3p\,\mathrm{d}^3q\,\mathcal W_{23}$.
With $a=|\mathbf p|$, $c=|\mathbf q|$, $\mu=\hat{\mathbf p}\cdot\hat{\mathbf z}$,
$\nu=\hat{\mathbf p}\cdot\hat{\mathbf q}$, and relative azimuth $\psi$ about
$\mathbf p$, the five remaining variables have bounds
\begin{equation}
 k\le a,c\le M/2,\qquad -1\le\mu,\nu\le1,\qquad 0\le\psi<2\pi,
\end{equation}
with measure $2\pi a^2c^2\mathrm{d} a\,\mathrm{d} c\,\mathrm{d}\mu\,\mathrm{d}\nu\,\mathrm{d}\psi$ after
the common azimuth is integrated. The exact $3k\le x\le M$, third-mode and
routing cuts are applied within these bounds. The upper bound $M/2$
for a single final momentum follows from the triangle inequality and
spatial closure. The probability density $g$ of the importance sampler
is expressed relative to $\mathrm{d}^3p\,\mathrm{d}^3q$ and averaged over six assignments
of the final labels. A sample contributes $\mathcal W_{23}/(3!\,g)$ if accepted.
The six-assignment average of $g$ and the final-state $1/3!$ have distinct
roles. The energy strata split at $x=16$; this split is not a physical cut.

\subsection{Retained amplitudes}
The numerical model is $S_1=\exp(i\delta T_1)$. The $2\to2$
result uses its leading coherent exchange contribution from
$-\tfrac12\delta T_1^2$ in the common normalization. Independent
higher-generator contact terms are outside this model; no contact
cancellation is assumed. The fusion routing through the exact spatial
zero mode does not contribute in the center of mass. The $2\to3$ amplitude used here is precisely
\begin{equation}
 A_{23}^{\mathrm{iter}}=-\frac{i}{3!}
 \langle f_3|(\delta T_1)^3|i_2\rangle_{\mathrm{conn}}.
 \label{eq:num-retained23}
\end{equation}
The cubic generator includes the splitting and joining vertices as a
Hermitian pair. The code sums the nonzero cubic routings and allowed
intermediate states coherently, with both internal helicities. Their
interference is included before taking the absolute square. The
projection uses generic distinct modes; disconnected spectator delta
terms are not included. The soft and near-spectator importance proposals
sample kinematics of the connected contribution, not additional
disconnected amplitudes.

Independent higher-generator terms are reserved for the extension in
Appendix~\ref{app:higher-generator-extension}; they are not part of the
present $S_1$ model. After cancellation of the common spatial
normalization, its leading channel weights are proportional to
$\varepsilon^4 I_{22}$ and $\varepsilon^6 I_{23}$. We set
$\varepsilon=1$ and use the unexpanded ratio of the retained weights.
This is a conditional mean of leading connected channels of $S_1$,
not its full inclusive expectation or a complete probability expansion
through order $\varepsilon^6$.

\subsection{Numerical accuracy and upper-energy dependence}
\paragraph{Sampling, refinement and stopping.}
The direct small-$k$ runs use eight independent scrambled Sobol replicas
per stratum. Each replica has 256 points below $x=16$ and 128 points in
the upper stratum, for 3072 five-dimensional samples. The $2\to2$ integral
uses deterministic Gauss--Legendre quadrature, of orders $(n_x,n_q)=(6,6)$,
in logarithmic energy and transfer coordinates. Panels resolve the folded
energy support changes and the scales $u-1\sim 2^j k$. Transfer panels
have endpoint ratios at most four. The angular kernel uses order 16
on its piecewise Gauss panels, with geometrically spaced polar subdivisions
that resolve the separation between the pole regions and the parent
energy cone. At angular order $m$, cap radial integrals have order $m$ and use $4m+8$
azimuthal midpoint nodes on the reflection half-domain. The cap and
complement integrals reconstruct the full sphere.

The direct integrations stop when all prescribed replica points,
deterministic nodes and selected checks have been evaluated. There is no
stopping rule based on a target error bar. The refinement is the
geometry-dependent placement of quadrature panels and the order-doubling
test, not an automatic increase of sample size until convergence.
The direct acceptance criteria are relative changes below $10^{-3}$
in checked nonzero weights at angular order $16\to32$, and below
$10^{-3}$ in $I_{22}$ for $n_x=4\to6$ at fixed $n_q=6$.
Angular test nodes are selected for large contributions to the
normalization or the first and second centered energy moments.
Physical rejection by the momentum and energy bounds yields a zero
sample contribution. A nonfinite accepted evaluation or an incomplete
replica is a failed calculation, not an extra zero contribution.

\paragraph{Covariance and deterministic quadrature.}
We estimate $D_{2n}=\int (x-1)\mathrm{d} I_{2n}$ directly and set
$T=I_{22}+I_{23}$, $D=D_{22}+D_{23}$.
The reported mean is $1+r_E$ with $r_E=D/T$.
For each independent energy stratum, the replica estimate is a pair
$(I_{23},D_{23})$, and its contribution to the covariance of the mean
is the sample covariance of the eight pairs divided by eight. The
covariances of independent strata are added. With $2\to2$ held fixed,
the propagated statistical variance is
\begin{equation}
 \operatorname{SE}^2(r_E)=\frac{\operatorname{Var}(D_{23})}{T^2}
 +\frac{D^2\operatorname{Var}(I_{23})}{T^4}
 -\frac{2D\operatorname{Cov}(I_{23},D_{23})}{T^3}.
 \label{eq:num-covariance}
\end{equation}
This uses the covariance of numerator and denominator from the same
samples. It is a statistical error of the estimated mean and differs
from the physical width of the outgoing-energy distribution.
At $k=10^{-8}$ and $M=2024$, the linear and leave-one-replica-out
errors are $4.6625\times10^{-8}$ and $4.6911\times10^{-8}$.
The deterministic $2\to2$ uncertainty is tested separately: the orders
$(4,6)$ and $(6,6)$ use 5730 and 8664 nodes at this point, change $I_{22}$
by $4.85\times10^{-6}$ relatively, and shift $r_E$ by
$1.99\times10^{-16}$ in absolute value. These checks do not provide
a global systematic error bound. The statistical errors do not cover
omitted amplitudes or a change of angular prescription.

\begin{table}[htbp]
\centering
\caption{Nested Sobol prefixes at fixed $k=10^{-8}$. Counts are per
replica, with eight replicas per stratum. The tail above $x=2024$
contributes only to the larger upper bound. Errors are statistical
standard errors. Prefix rows are correlated.}
\label{tab:num-prefix}
\begin{tabular}{rrrcc}
\toprule
$N_{<16}$ & $N_{16\text{--}2024}$ & $N_{>2024}$ &
$10^7r_E$, $M=2024$ & $10^7r_E$, $M=18096$\\
\midrule
64 & 32 & 32 & $5.107\pm0.750$ & $5.681\pm0.810$\\
128 & 64 & 64 & $6.068\pm0.295$ & $6.607\pm0.330$\\
256 & 128 & 128 & $6.114\pm0.466$ & $6.612\pm0.494$\\
\bottomrule
\end{tabular}
\end{table}

\begin{table}[htbp]
\centering
\caption{Direct angular and radial checks at $M=2024$. The angular
column is the largest observed relative weight change among the
listed pointwise comparisons, not an error bound on the full integral.}
\label{tab:num-angular-checks}
\begin{tabular}{r r c c c}
\toprule
$k$ & Angular nodes & $\max|P_{32}/P_{16}-1|$ &
$|I_{22,6}/I_{22,4}-1|$ & $|\Delta r_E|$ (radial)\\
\midrule
$10^{-8}$ & 41 & $1.36\times10^{-5}$ & $4.85\times10^{-6}$ & $1.99\times10^{-16}$\\
$10^{-7}$ & 26 & $1.31\times10^{-5}$ & $4.85\times10^{-6}$ & $1.60\times10^{-14}$\\
$10^{-6}$ & 26 & $1.35\times10^{-5}$ & $4.85\times10^{-6}$ & $1.26\times10^{-12}$\\
\bottomrule
\end{tabular}
\end{table}

The last doubling in Table~\ref{tab:num-prefix} changes $r_E$ at $M=2024$
by $4.63\times10^{-9}$, about $0.10$ of the final statistical error.
The errors of the different prefixes need not decrease monotonically.
Table~\ref{tab:num-angular-checks} reports the observed angular and radial
changes. In a separate paired calculation at $k=10^{-7}$ and $M=2048$,
halving the radius fraction from $0.35$ to $0.175$ shifted the mean by
$2.29\times10^{-13}$, compared with a standard error of
$2.71\times10^{-7}$; 82 order-doubling checks gave a maximum relative
weight change of $1.37\times10^{-5}$. This tests the integration
partition within the same circular prescription.

\paragraph{Completed upper-energy scan at $k=10^{-8}$.}
We extend the verified $M=2024$ calculation by sampling the disjoint
tail $2024<x\le18096$. This uses eight additional independent scrambled
Sobol replicas of 128 points at angular order 16. The energetic leg in
the tail proposal has $c\ge2024/3$; Bose symmetrization preserves support
since at least one leg has energy at least $x/3$. The same tail points and
their proposal weights are used at all larger $M$, with the indicator
$\mathbf 1(x\le M)$. The original $x\le2024$ estimates are common to all
rows. The $2\to2$ quadrature is evaluated for each upper bound, reusing
integrand values only at identical nodes. This gives a paired comparison
of the bounds, rather than independent estimates of nearly equal means.

\begin{table}[htbp]
\centering
\small
\caption{Upper-energy scan at $\kappa_{\mathrm{IR}}/E_i=10^{-8}$.
The mean is $1+r_E$, and $p_{23}=1-p_{22}$ with the same standard error.
The differences are measured relative to $M=2024$, with the full
cross-bound statistical covariance. Errors are one standard error.}
\label{tab:num-emax}
\begin{tabular}{r c c c c}
\toprule
$M$ & $p_{22}$ (\%) & $p_{23}$ (\%) & $10^7r_E$ & $10^7[r_E(M)-r_E(2024)]$\\
\midrule
2024 & $0.00670\pm0.00043$ & $99.99330\pm0.00043$ & $6.114\pm0.466$ & $0$ (reference)\\
4048 & $0.00670\pm0.00043$ & $99.99330\pm0.00043$ & $6.229\pm0.472$ & $+0.114\pm0.013$\\
16192 & $0.00670\pm0.00043$ & $99.99330\pm0.00043$ & $6.592\pm0.493$ & $+0.478\pm0.042$\\
18096 & $0.00670\pm0.00043$ & $99.99330\pm0.00043$ & $6.612\pm0.494$ & $+0.498\pm0.044$\\
\bottomrule
\end{tabular}
\end{table}

The three larger bounds yield 243, 762, and 802 accepted tail points
from 1024 common proposals. In 28 order-doubling checks the maximum
relative weight change is $1.03\times10^{-5}$; radial refinement
changes $I_{22}$ by at most $4.85\times10^{-6}$ relatively and $r_E$
by $2.15\times10^{-16}$ absolutely. From $M=2024$ to 18096, the
paired increase is $(0.498\pm0.044)\times10^{-7}$; between the two
largest bounds it is $(1.98\pm0.58)\times10^{-9}$. The bounds
16192 and 18096 correspond to $4\times4048$ and $2\times9048$.
The earlier rounded local estimate at $M=4048$ was
$10^7r_E=6.69\pm0.60$, whereas Table~\ref{tab:num-emax} reports
the paired estimate. Because the differences use common-sample
covariances, they cannot be reconstructed from marginal errors alone.
These finite-cutoff results do not establish the $M\to\infty$ limit.

\paragraph{Archived infrared profiles.}
Sample counts for the direct low-cutoff runs do not apply to all plotted
points. Some archived runs used the stochastic angular correction
\begin{equation}
 \widehat{\mathbf m}=\frac{1}{N}\sum_i\mathbf m_i^{\rm base}
 +\frac{1}{N}\sum_{i\in S}
 \frac{\mathbf m_i^{\rm fine}-\mathbf m_i^{\rm base}}{\pi_i},
\end{equation}
with $\pi_i$ the inclusion probability of a refined point. Each
archived run used four replicas. At $k=10^{-5}$, each replica had
256 low-energy and 128 upper-energy points, with 89 order-32
corrections; the largest $2\to3$ angular validation change was
$3.92\times10^{-7}$. At $k=5\times10^{-4}$, the corresponding
counts were 512 and 128, with 157 order-64 corrections. Six runs for
$10^{-3}\le k\le0.05$ used 256 and 64 points, with 85--101
order-64 corrections; the largest of six selected order-128 versus
order-64 weight changes was $5.49\times10^{-4}$. Runs at
$k\simeq2.66\times10^{-5},7.07\times10^{-5},1.88\times10^{-4}$
used 256 and 128 points, with 103, 96, and 105 order-32 corrections,
respectively. Rounded table entries alone do not certify convergence
or sample counts for other local runs.

\paragraph{Complete numerical results.}
Table~\ref{tab:nsf-ir-channel-weights-full} records the full set of
published table entries for the conditional mean energy and both channel
weights. Table~\ref{tab:nsf-ir-channel-weights} in the main text selects
the points needed to discuss the maximum, the decrease of the residual,
and the growing relative weight of the $2\to3$ channel.

\begin{table}[t]
  \centering
  \caption{Conditional mean outgoing energy and relative weights of the retained $2\to2$ and $2\to3$
  processes as functions of the infrared cutoff, at
  $E_i=G=\epsilon=1$ and $E_{\max}=2024E_i$.
  The weights are normalized within these two channels:
  $p_{22}=I_{22}/(I_{22}+I_{23})$ and
  $p_{23}=I_{23}/(I_{22}+I_{23})$.
  The channel weights are reported as percentages.}
  \label{tab:nsf-ir-channel-weights-full}
    \begin{tabular}{cccc}
    \toprule
    $\kappa_{\mathrm{IR}}/E_i$ & $p_{22}\;(\%)$ & $p_{23}\;(\%)$ &
    $\overline{E}_{\mathrm{out}}/E_i$ \\
    \midrule
    $10^{-8}$ & $0.00670 \pm 0.00043$ & $99.99330 \pm 0.00043$ & $1.000000611 \pm 0.000000047$ \\
    $10^{-7}$ & $0.0629 \pm 0.0043$ & $99.9371 \pm 0.0043$ & $1.00000524 \pm 0.00000027$ \\
    $10^{-6}$ & $0.595 \pm 0.034$ & $99.405 \pm 0.034$ & $1.0000438 \pm 0.0000028$ \\
    $10^{-5}$ & $5.87 \pm 0.88$ & $94.13 \pm 0.88$ & $1.000357 \pm 0.000062$ \\
    $5\times10^{-4}$ & $77.0 \pm 1.4$ & $23.0 \pm 1.4$ & $1.00318 \pm 0.00026$ \\
    $10^{-3}$ & $88.7 \pm 1.0$ & $11.3 \pm 1.0$ & $1.00294 \pm 0.00014$ \\
    $5\times10^{-3}$ & $96.71 \pm 0.15$ & $3.29 \pm 0.15$ & $1.00131 \pm 0.00021$ \\
    $10^{-2}$ & $98.065 \pm 0.067$ & $1.935 \pm 0.067$ & $0.99866 \pm 0.00014$ \\
    $2\times10^{-2}$ & $99.025 \pm 0.044$ & $0.975 \pm 0.044$ & $0.992884 \pm 0.000069$ \\
    $5\times10^{-2}$ & $99.566 \pm 0.039$ & $0.434 \pm 0.039$ & $0.977483 \pm 0.000093$ \\
    \bottomrule
  \end{tabular}  \par\smallskip
  \begin{minipage}{0.95\linewidth}
    \footnotesize
    Quoted uncertainties are one statistical standard error. The four
    lowest-cutoff entries correspond to the reference infrared scan
    described in Appendix~\ref{app:numerical-implementation}.
  \end{minipage}
\end{table}

\begin{figure}[tbp]
\centering
\includegraphics[width=0.85\linewidth]{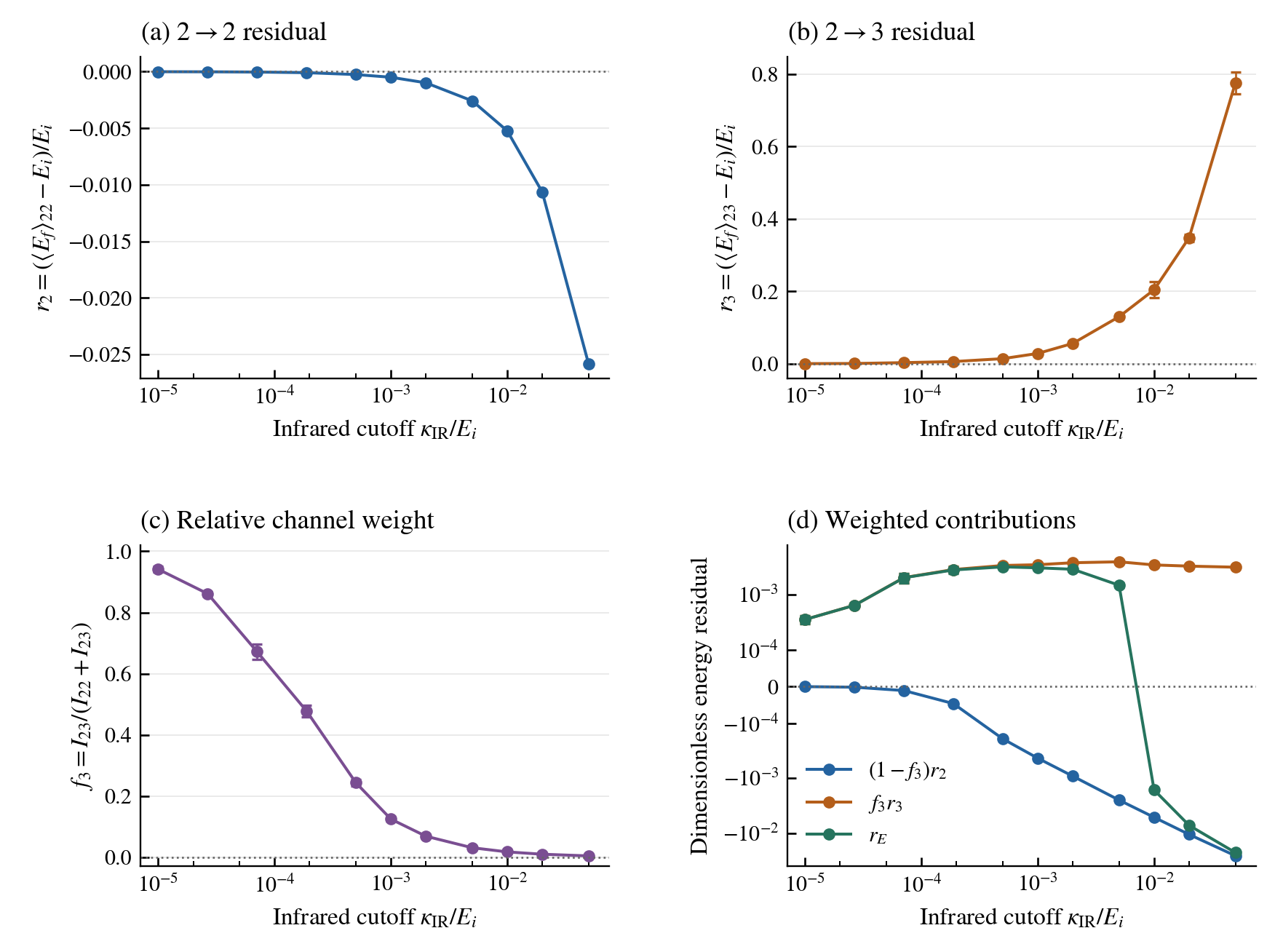}
 \caption{Infrared-cutoff dependence of the conditional outgoing-energy
  residuals and relative channel weights for the retained NSF amplitudes.
  The retained amplitudes are the $2\to2$ cubic-exchange contribution
  and $\mathcal A_{23}^{\rm iter}$ of
  Eq.~\eqref{eq:numerical-retained23}; all retained routings interfere
  coherently. Both are leading connected amplitudes of
  $S_1=\exp(i\delta T_1)$. Panels (a) and (b) show
  $r_2=\langle E_f\rangle_{22}/E_i-1$ and
  $r_3=\langle E_f\rangle_{23}/E_i-1$, respectively, where $E_f$ is the
  total outgoing energy. Panel (c) shows
  $f_3=I_{23}/(I_{22}+I_{23})$. Panel (d) resolves the combined residual
  $r_E=(1-f_3)r_2+f_3r_3$ into its two channel contributions.
  The calculation uses $E_i=G=\epsilon=1$ and
  $E_{\max}=2024E_i$, without imposing an external energy-conserving
  delta function. Error bars denote one statistical standard error
  propagated with numerator--denominator covariance from the $2\to3$
  replicas, with deterministic $2\to2$ held fixed. The documented direct
  low-cutoff runs use eight replicas per stratum, whereas the $10^{-5}$
  archive uses four. Angular-prescription and omitted-amplitude
  uncertainties are not included; numerical checks appear in
  Appendix~\ref{app:numerical-implementation}.
  All plotted points use the four infrared cutoffs
  $\kappa_{\rm IR}/E_i=10^{-5},10^{-6},10^{-7},10^{-8}$
  and the retained calculation used for the tabulated results, with
  the run-specific sampling profiles documented in
  Appendix~\ref{app:numerical-implementation}.
  Lines connecting the evaluated points guide the eye.
  Panel (d) uses a symmetric logarithmic vertical scale, linear for
  $|r|\leq10^{-4}$.}
 \label{fig:unreduced-energy-residual}
\end{figure}

\paragraph{Channel-resolved energy residuals.}
The diagnostic plots in Fig.~\ref{fig:unreduced-energy-residual}
separate the two channel means and display their contributions to the
combined residual; they are not additional assumptions or independent
energy-conservation tests.

\section{Outgoing-energy variance in the retained channels}
\label{app:energy-variance}

The first energy moment alone does not determine whether the outgoing
energy distribution concentrates at $E_i$. We therefore also consider
its second centered moment, using the same reduced amplitudes,
integration domain, helicity sums, Bose factors, and circular finite-part
prescription as in Appendix~\ref{app:connected-channel-energies}.
No additional amplitudes are included in this test. Define $X=E_f/E_i$
and, for $n=2,3$,
\begin{align}
 Q_n&=\mathcal J_n[E_f^2|\mathcal A_{G,2n}|^2],&
 C_n&=\mathcal J_n[(X-1)^2|\mathcal A_{G,2n}|^2].
 \label{eq:variance-new-moments}
\end{align}
Here $P_n$ and $W_n$ retain the definitions in
Eqs.~\eqref{eq:appendix-channel22-moments} and
\eqref{eq:appendix-channel23-moments}. With
$P=P_2+P_3$, $W=W_2+W_3$, $Q=Q_2+Q_3$ and $C=C_2+C_3$,
\begin{equation}
 \mu=\frac{W}{E_iP}=1+r_E,\qquad
 v\equiv\operatorname{Var}(X)
 =\frac{C}{P}-r_E^2
 =\frac{Q}{E_i^2P}-\mu^2.
 \label{eq:variance-main-identity}
\end{equation}
The dimensional variance is $E_i^2v$ and the physical standard
deviation is $E_i\sqrt v$. The integration imposes no delta of
external energy conservation.

The channel means and variances are
$\mu_n=W_n/(E_iP_n)$ and
$v_n=Q_n/(E_i^2P_n)-\mu_n^2$; writing $p_n=P_n/P$ gives the
independent consistency identity
\begin{equation}
 v=p_2v_2+p_3v_3+p_2p_3(\mu_2-\mu_3)^2.
 \label{eq:variance-total-law}
\end{equation}
The relative channel weights and total means are already tabulated in
Table~\ref{tab:nsf-ir-channel-weights-full} and are not repeated here.

\begin{table}[htbp]
\centering\small
\caption{Physical variance of $X=E_f/E_i$ for the retained
$2\to2$ and $2\to3$ channels at $E_i=G=\varepsilon=1$ and
$E_{\max}=2024E_i$. The errors on $v$ and $\sqrt v$ are
statistical standard errors of the estimators, not part of the
physical energy spread. Numerical values are those supplied by the
channel-moment analysis.}
\label{tab:variance-ir}
\begin{tabular}{ccccc}
\toprule
$\kappa_{\rm IR}/E_i$ & $v_2$ & $v_3$ & $v\pm\operatorname{SE}(\widehat v)$
& $\sqrt v\pm\operatorname{SE}(\widehat{\sqrt v})$\\
\midrule
$10^{-5}$ & $9.396\times10^{-6}$ & $3.230\times10^{-2}$ & $(3.040\pm0.784)\times10^{-2}$ & $(1.744\pm0.225)\times10^{-1}$\\
$10^{-6}$ & $9.390\times10^{-7}$ & $3.882\times10^{-3}$ & $(3.859\pm0.351)\times10^{-3}$ & $(6.212\pm0.283)\times10^{-2}$\\
$10^{-7}$ & $9.389\times10^{-8}$ & $4.060\times10^{-4}$ & $(4.057\pm0.324)\times10^{-4}$ & $(2.014\pm0.081)\times10^{-2}$\\
$10^{-8}$ & $9.389\times10^{-9}$ & $4.021\times10^{-5}$ & $(4.020\pm0.337)\times10^{-5}$ & $(6.341\pm0.266)\times10^{-3}$\\
\bottomrule
\end{tabular}
\end{table}

At fixed $E_{\max}$ the measured variance decreases together with the
mean-energy residual as the infrared cutoff is reduced. The variance
is predominantly the within-channel term $p_2v_2+p_3v_3$; the
between-channel term in Eq.~\eqref{eq:variance-total-law} is at most
$8.15\times10^{-9}$ over the four tabulated cutoffs. Positivity
$C/P\ge r_E^2$ was checked at every point. The reported numerical
cross-checks give a maximum absolute difference
$5.633\times10^{-16}$ between the raw- and centered-moment formulas
and $4.337\times10^{-19}$ in the total-variance identity.
These are algebraic consistency checks on the supplied numerical
moments, not an independent validation of the underlying angular
prescription.

\paragraph{Estimators and statistical resolution.}
For numerical stability the centered first moment
$D=W/E_i-P$ is accumulated together with $P$ and $C$. Every accepted
phase-space sample with importance weight $w$ contributes
$w$, $w(X-1)$ and $w(X-1)^2$ respectively. The distinct folded
$2\to2$ energy branches are separately evaluated. For the older
$10^{-5}$ sample, the documented inclusion-probability correction
for angular reevaluations must be applied to all three moments.
Independent scrambled Sobol replicas yield the covariance of the
estimated moment vector, with independent energy-stratum covariances
added. If $B=W/E_i$ and $\Sigma$ is the covariance matrix of
$(P,B,C)$, the delta-method statistical standard error is
\begin{equation}
 \operatorname{SE}(\widehat v)=\sqrt{g^{\mathsf T}\Sigma g},\qquad
 g=\left(-\frac{C}{P^2}+\frac{2r_EB}{P^2},
 -\frac{2r_E}{P},\frac1P\right),\qquad
 \operatorname{SE}(\widehat{\sqrt v})\simeq
 \frac{\operatorname{SE}(\widehat v)}{2\sqrt v}.
 \label{eq:variance-error}
\end{equation}
This statistical uncertainty is distinct from the physical variance
$v$. As in Appendix~\ref{app:numerical-implementation}, the
$2\to2$ quadrature is held fixed in this covariance and its numerical
resolution is checked separately. The $10^{-5}$ calculation uses four
replicas (1536 basic points); each lower-cutoff calculation uses eight
replicas (3072 points). The reported leave-one-replica-out errors
differ by no more than $3.14\%$ from the linear errors. At radial
order $(n_x,n_q)=(6,6)$ versus $(4,6)$, the largest observed shift
in $v$ is $8.639\times10^{-9}$ at $10^{-5}$ and decreases at smaller
cutoffs. These checks do not establish a global systematic error bound.

\paragraph{Upper-energy sensitivity.}
The corresponding second-moment scan at $\kappa_{\rm IR}/E_i=10^{-8}$,
using the paired upper-bound protocol of Table~\ref{tab:num-emax}, is
\begin{table}[htbp]
\centering\small
\caption{Outgoing-energy variance at fixed infrared cutoff
$\kappa_{\rm IR}/E_i=10^{-8}$, showing sensitivity to the upper-energy
cut. Errors are statistical standard errors; $r_E$ is repeated only
to compare first and second moments.}
\label{tab:variance-upper-bound}
\begin{tabular}{rccc}
\toprule
$E_{\max}/E_i$ & $v\pm\operatorname{SE}(\widehat v)$ &
$\sqrt v\pm\operatorname{SE}(\widehat{\sqrt v})$ & $r_E$\\
\midrule
2024 & $(4.020\pm0.337)\times10^{-5}$ & $(6.341\pm0.266)\times10^{-3}$ & $6.1144\times10^{-7}$\\
4048 & $(7.438\pm0.603)\times10^{-5}$ & $(8.624\pm0.349)\times10^{-3}$ & $6.2285\times10^{-7}$\\
16192 & $(3.768\pm0.323)\times10^{-4}$ & $(1.941\pm0.083)\times10^{-2}$ & $6.5924\times10^{-7}$\\
18096 & $(4.106\pm0.352)\times10^{-4}$ & $(2.026\pm0.087)\times10^{-2}$ & $6.6122\times10^{-7}$\\
\bottomrule
\end{tabular}
\end{table}
The variance increases by a factor of approximately $10.21$ between
the smallest and largest upper bounds, with paired difference
$\Delta v=(3.704\pm0.333)\times10^{-4}$, whereas the first moment
changes far less. The variance is therefore not ultraviolet-cutoff
independent over the tested domain. No joint infrared--ultraviolet
limit is inferred.

A separate paired test at $\kappa_{\rm IR}/E_i=10^{-7}$ and
$E_{\max}=2048E_i$ changes the angular partition radius fraction
from $0.35$ to $0.175$, holding the full sphere fixed. It gives
$v_{0.35}=4.104856069\times10^{-4}$ and
$v_{0.175}=4.104857198\times10^{-4}$, a difference
$\Delta v=1.130\times10^{-10}$ with paired statistical error
$1.317\times10^{-11}$. This tests the angular partition within the
circular prescription, not equivalence to an alternative treatment
of the singular kernel.

The reported decrease of both $|r_E|$ and $v$ with decreasing
$\kappa_{\rm IR}$ at fixed $E_{\max}$ is evidence of an increasingly
narrow conditional energy distribution within the retained model.
It does not prove exclusive energy conservation, the full inclusive
balance, or $[H,S]=0$. The uncertainties do not cover omitted
higher-generator amplitudes, angular-prescription systematics, or an
uncontrolled energy-cutoff limit.

\section{Zero-frequency News and the shear memory}\label{app:news-memory}

At each angle, use the Fourier convention
\begin{equation}
 \widetilde f(\omega)
 =\int_{-\infty}^{+\infty}du\,e^{-i\omega u}f(u).
\end{equation}
Then, in the sense of tempered distributions,
\begin{equation}
 \widetilde N^+(\omega,\Omega)
 =i\omega\,\widetilde\sigma^+(\omega,\Omega),
 \qquad N^+=\dot\sigma^+.
 \label{eq:Fourier-News-shear}
\end{equation}
If $N^+$ is integrable in $u$ and the shear has finite endpoint limits,
continuity of its Fourier transform at the origin yields
\begin{equation}
 \widetilde N^+(0,\Omega)
 =\int_{-\infty}^{+\infty}du\,N^+(u,\Omega)
 =\sigma^+(+\infty,\Omega)-\sigma^+(-\infty,\Omega)
 \equiv\Delta\sigma^+(\Omega)=-\sigma_0^+(\Omega).
 \label{eq:news-zero-memory}
\end{equation}
If the non-$\delta$ part of $\widetilde\sigma^+$ is regular and bounded near
$\omega=0$, then
Eq.~\eqref{eq:Fourier-News-shear} implies
$\widetilde N^+(0)=0$ and hence $\Delta\sigma^+=0$.
Conversely, a nonzero memory gives
$\widetilde N^+(\omega)\to\Delta\sigma^+$ and forces a distributional
$1/\omega$ behavior in the shear:
\begin{equation}
 \widetilde\sigma^+(\omega,\Omega)
 \sim \frac{\Delta\sigma^+(\Omega)}{i}\,
       \operatorname{PV}\!\left(\frac{1}{\omega}\right)
 \qquad (\omega\to0),
 \label{eq:shear-soft-pole}
\end{equation}
up to a term proportional to $\delta(\omega)$, which records the choice of
constant shear offset, and terms less singular at the origin.  This is the
infrared memory sector.  The condition $\Delta\sigma=0$ characterizes the
\emph{memory-free} radiative subsector; configurations with
$\Delta\sigma\neq0$ can still carry ordinary finite-frequency News and
positive News energy flux.  In the DKL expression for $\chi$, their soft
shear contribution is $\bar\eth^{\,2}\sigma_0^+$, but it continues to
have zero projection on the four translation harmonics
\cite{DominguezKozamehLudvigsen1996,DominguezKozamehLudvigsen1997,StromingerZhiboedov2016}.
The density of $\mathcal D_{\rm rad}$ in the ordinary finite-frequency
Fock norm does not extend the condition $\Delta\sigma=0$ to configurations
with memory: the zero-frequency functional in
Eq.~\eqref{eq:news-zero-memory} is not controlled merely by that norm.
Thus the weak-to-operator argument on $\mathcal D_{\rm rad}$ concerns the
chosen radiative representation, without erasing physically relevant
memory sectors.

\end{document}